\documentclass[12pt, peerreview , onecolumn]{IEEEtran}

\usepackage{tikz}
\usepackage[sumlimits,intlimits]{amsmath}
\usepackage{amsmath,amsfonts,amssymb}%
\usepackage{bm}%
\usepackage{graphicx,graphics}%
\usepackage{epstopdf}
\usepackage{cases}%
\usepackage[noadjust]{cite}%
\usepackage{color}%
\usepackage{cite,url}
\usepackage{verbatim}
\usepackage{algorithm}
\usepackage{algorithmic}
\usepackage{multirow}
\usepackage{multicol}
\usepackage{stfloats}
\usepackage{subfigure}
\usepackage{xspace}
\usepackage{enumerate}
\usepackage{rotating}
\usepackage[numbers,sort&compress]{natbib}

\ifCLASSOPTIONpeerreview

\fi

\begin{document}

\ifCLASSOPTIONpeerreview
\title{\hspace*{-12pt}{Joint Data Detection and Phase Noise Mitigation \\ \vspace*{-10pt}
    for Light Field Video Transmission \\ \vspace*{-10pt}
    in MIMO-OFDM Systems}}
\fi

\author{\IEEEauthorblockN{Omar H. Salim, , \textit{Member, IEEE},  Wei Xiang, \textit{Senior Member, IEEE}, Ali A.~Nasi, \textit{Member, IEEE}}, Gengkun Wang, and Hani Mehrpouyan, \textit{Senior Member, IEEE}}

\maketitle

{\let\thefootnote\relax\footnotetext{ Omar H. Salim and Gengkun Wang are with the School of Mechanical and Electrical Engineering, University of Southern Queensland, Australia.  Wei Xiang is with the Department of Electronic Systems and Internet of Things Engineering, James Cook University, Australia. Ali A. Nasir is with School of Electrical Engineering and Computer Science, National University of Sciences and Technology, Islamabad, Pakistan. Hani Mehrpouyan is with the Department of Electrical and Computer Engineering, Boise State University. Emails: OmarHazim.Salim@usq.edu.au, wei.xiang@jcu.edu.au, ali.nasir@seecs.edu.pk,  Gengkun.Wang@usq.edu.au, and hani.mehr@ieee.org. \vspace{-00pt}}}

\vspace{-12pt}

\begin{abstract}
Previous studies in the literature for video transmission over wireless communication systems focused on combating the effects of additive channel noise and fading channels without taking the impairments in the physical layer such as phase noise (PHN) into account. Oscillator phase noise impairs the performance of multi-input multi-output-orthogonal frequency division multiplexing (MIMO-OFDM) systems in providing high data rates for video applications and may lead to decoding failure. In this paper,  we propose a light field (LF) video transmission system in wireless channels, and analyze joint data detection and phase mitigation in MIMO-OFDM systems for LF video transmission. The signal model and rate-distortion (RD) model for LF video transmission in the presence of multiple PHNs are discussed. Moreover, we propose an iterative algorithm based on the extended Kalman filter for joint data detection and PHN tracking. Numerical results show that the proposed detector can significantly improve the average bit-error rate (BER) and peak-to-noise ratio (PSNR) performance for LF video transmission compared to existing algorithms. Moreover, the BER and PSNR performance of the proposed system is closer to that of the ideal case of perfect PHN estimation. Finally, it is demonstrated that the proposed system model and algorithm are well suited for LF video transmission in wireless channels. 
\end{abstract}

\ifCLASSOPTIONpeerreview
    \newpage
\fi

\section{Introduction}
\subsection{Motivation and Literature Survey}
The next generation (5G) communication networks are expected to provide three-dimensional (3-D) video services with high data rate in gigabits per second (Gbps), which can be achieved through adopting multi-input multi-output (MIMO) systems and by using the large bandwidth in the millimeter-wave band, e.g., V-band/$60$ GHz and E-band/$70$-$80$ GHz \cite{{Xiang:2015Jan},{article_mehr-comm-13}}. Recently, light field (LF) technology is emerged with the advance of computational photography to solve several challenges in computer vision such as virtual refocusing and synthetic aperture, and rendering images from a virtual viewpoint \cite{Yang:2008LF}. Both special and angular information can be captured by LF photography. Thus, professional photographers adopt LF photography to create animated and multi-focus digital photos. In addition, the depth perception in LF technology enables LF to be implemented in a wide range of industry applications \cite{{Xiang:2015Jan},{Xiang:2015Oct},{Xiang:2015crossLF}}. 
However, the delivery of 3-D video services over MIMO systems poses new challenges. This is due to oscillator fluctuations, which affect the transmitted signal with different aspects such as the phase noise (PHN) and carrier frequency offset (CFO). Moreover, PHN has a greater impact on the system performance at higher carrier frequencies \cite{article_mehr-comm-13}. In contrast to the CFO, which can be estimated and compensated during the transmission of a training sequence, PHN is time-varying in nature and needs to be tracked not only during the training interval but also during the data transmission interval \cite{{Lin:2007May},{Hani:2012sept},{Nasir:2013Jan}}. Moreover, PHN can significantly reduce the channel capacity of wireless communication systems and limit their performance in providing video services with high quality \cite{{Mathecken:2011May},{Mathecken:2012Sep}}. Therefore, it is increasingly important to develop efficient and accurate estimation algorithms to enhance the robustness of wireless communication systems and overcome the effect of PHN to provide satisfying services for LF video applications \cite{LANMAN:2010ACM}. 

PHN estimation in MIMO systems is more challenging than in signal-input single-output (SISO) systems, since MIMO systems have multiple antennas and each one may have its own local oscillator \cite{Hani:2012sept}. Thus, this gives rise to multiple PHNs \cite{{Hani:2012sept},{Nasir:2013Jan}}. In addition, data detection for each pair of antennas becomes challenging in the presence of PHN parameters \cite{{Lin:2007May},{Hani:2012sept},{Nasir:2013Jan}}. Therefore, video transmission is only possible if the PHN parameters can be estimated at the receiver \cite{Nasir:2011Dec}.

Orthogonal frequency division multiplexing (OFDM) systems are employed in MIMO systems to increase spectral efficiency and mitigate the effect of frequency-selective fading. However, OFDM systems are much
more sensitive to synchronization errors caused by PHN, since PHN is convolved with the received symbols, which results in a common phase error (CPE) and inter-carrier interference (ICI)  \cite{{Armada:2001Jun},{Petrovic:ICC2005May},{Petrovic:2007Aug}}. The CPE is a multiplicative factor that affects all subcarriers similarly \cite{{Petrovic:ICC2005May},{Petrovic:2007Aug}}, while
the ICI is an additive PHN factor \cite{{Wu:2002Dec},{Schenk:2004Sep}}.

In \cite{Wang:2010WCNCJul}, a new algorithm for estimation of the CPE in space division multiple access (SDMA) MIMO-OFDM systems was proposed. Even though the maximum a posteriori estimator in \cite{Wang:2010WCNCJul} can track multiple PHN parameters, it has a very high
computational complexity, and the performance of proposed estimators is only verified for low-to-moderate phase noise variances. In addition, the ICI mitigation is ignored in \cite{Wang:2010WCNCJul}. The results in \cite{Schenk:2005Aug} provide schemes for mitigating the effect of PHN induced ICI in MIMO-OFDM systems. However, they present no means of estimating the CPE or the multiplicative PHN affecting these communication systems. Moreover, no data detection is presented in \cite{Schenk:2005Aug}. Algorithms for CPE estimation in MIMO-OFDM systems were proposed in \cite{{Schenk:2005Aug},{Liu:2006Mar},{Bittner:2008Mar}}, which are limited to scenarios where a single oscillator is used at all transmit or receive antennas. As a result, the approaches in \cite{{Schenk:2005Aug},{Liu:2006Mar},{Bittner:2008Mar}} cannot track multiple PHN parameters and cannot be applied to various MIMO systems. Iterative data detection based on the expectation-conditional maximization (ECM) algorithm was proposed in \cite{Pun:2007Feb} for orthogonal frequency division multiple access (OFDMA) uplink systems. However, the authors did not take the effect of PHN into account. In \cite{Hani:2012sept}, the authors proposed a decision-directed extended Kalman filter for tracking the PHN throughout a data packet in a \textit{single} carrier MIMO systems. 
However, the proposed method in \cite{Hani:2012sept} cannot be applied for multi-carrier OFDM systems, since the discrete Fourier transform (DFT) of PHN symbols are circularly convolved with the data and each data symbol is affected by the CPE and ICI after DFT operation in OFDM systems.
Recently, joint data detection and PHN mitigation are analyzed for cooperative systems in \cite{{Rabiei:2011May},{Wang:arxiv}}. However, the PHN tracking in \cite{{Rabiei:2011May},{Wang:arxiv}} requires the application of pilots throughout an OFDM data symbol to compensate the CPE, which adversely affects the bandwidth efficiency and data detection performance. Furthermore, the data detection approach using pilots to track the PHN over the data packet, as used in \cite{{Rabiei:2011May},{Wang:arxiv}}, has poor bit-error-rate (BER) and lower PHN estimation performance compared to the extended Kalman filter (EKF) based proposed in this paper.

In the context of video systems, the MIMO and MIMO-OFDM systems were proposed to enhance video transmission in \cite{{Zheng:2007Jan},{Song:2007Sep},{Xiao:2010July},{Salim:2011DICTA},{Salim:2012AugJN},{Liu:SPJN2014Jan},{Salim:2013Apr}}. However, the proposed systems for video transmission in \cite{{Zheng:2007Jan},{Song:2007Sep},{Xiao:2010July},{Salim:2011DICTA},{Salim:2012AugJN},{Liu:SPJN2014Jan},{Salim:2013Apr}} are based on the assumption that the PHN parameters are perfectly known at the receiver. In \cite{{Alajel:2011Dec},{Alajel:2012Oct}}, a relay selection method was proposed to improve the transmission of 3-D video over cooperative systems. However, in \cite{{Alajel:2011Dec},{Alajel:2012Oct}}, the effect of PHNs on data detection were not taken into account. In general, previous studies in the literature for video transmission over wireless communication systems have focused on combating the effects of additive noise and fading channels without taking the hardware impairments in the physical layer into account. 

\subsection{Contributions}
To the best of our knowledge, a complete analysis of  joint data detection and multiple PHN parameter estimation in MIMO-OFDM systems has not been addressed in the literature to date. In addition, the estimation effects of multiple PHN parameters on video transmission in MIMO-OFDM systems has not been studied in the existing literature. Moreover, there is no published work in LF video transmission over wireless channels. In view of these facts, joint data detection and multiple PHN parameter estimation in a MIMO-OFDM system equipped with $N_t$ transmit and $N_r$ receive antennas are analyzed, and the effects of PHN estimation on LF video applications over a MIMO-OFDM system are also investigated. The contributions of this paper can be summarized as follows:
\begin{itemize}
	\item An iterative algorithm based on the EKF for data detection and tracking the unknown time-varying PHN throughout the OFDM data packet in an $N_t \times N_r$ MIMO-OFDM system is proposed. Simulations are carried out to investigate the performance of the proposed detector. The simulation results demonstrate that the iterative receiver outperforms existing algorithms in \cite{{Rabiei:2011May},{Wang:arxiv}} in terms of bit error rate (BER) performance at different signal-to-noise ratios (SNRs). In addition, the BER performance of the proposed system is closer to that of scenario where no PHN is present.
	\item The effects of PHN on LF video applications over MIMO-OFDM systems are investigated. It is indicated that the proposed detector can significantly improve the peak signal-to-noise ratio (PSNR) with a variety of PHN variances. Simulation results also show that the iterative algorithm proposed in this paper can significantly outperform the existing estimation and conventional approaches for video transmission.
\end{itemize}

\subsection{Organization}
The rest of this paper is organized as follows: Section II describes the system model, Section III presents the problem formulation. Section IV derives the proposed detector. Section V illustrates the complexity analysis for the proposed system and Section VI provides simulation results and discussions. Finally, Section VII concludes the paper and summarizes its key findings.

\textit{Notation}

Superscripts $(\cdot)^\ast$, $(\cdot)^H$, and $(\cdot)^T$ denote the conjugate, the conjugate transpose, and the transpose operators, respectively. Bold face small letters, e.g., $\mathbf{x}$, are used for vectors in the time domain, bold face capital alphabets, e.g., $\mathbf{X}$, are used for matrices and vectors in the frequency domain, and $[\mathbf{X}]_{x,y}$ represents the entry in row \emph{x} and column \emph{y} of $\mathbf{X}$. $\mathbf{I}_X$, $\mathbf{0}_{X \times X}$, and $\mathbf{1}_{X \times X}$ denote the $X \times X$ identity, all zero, and all 1 matrices, respectively. \text{diag}($\mathbf{x}$) is used to denote a diagonal matrix, where the diagonal elements are given by vector $\mathbf{x}$. $\mathcal{N}(\mu,\sigma^2)$ and $\mathcal{CN}(\mu,\sigma^2)$ denote real and complex Gaussian
distributions with mean $\mu$ and variance $\sigma^2$, respectively. $\odot$ denotes circular convolution. Finally, $\dot{z}$ denotes the Jacobian of $z$.

\section{System Model} \label{sec:sigmodl}
\begin{figure}[htbp]
	\begin{center}
		\includegraphics[scale=0.4]{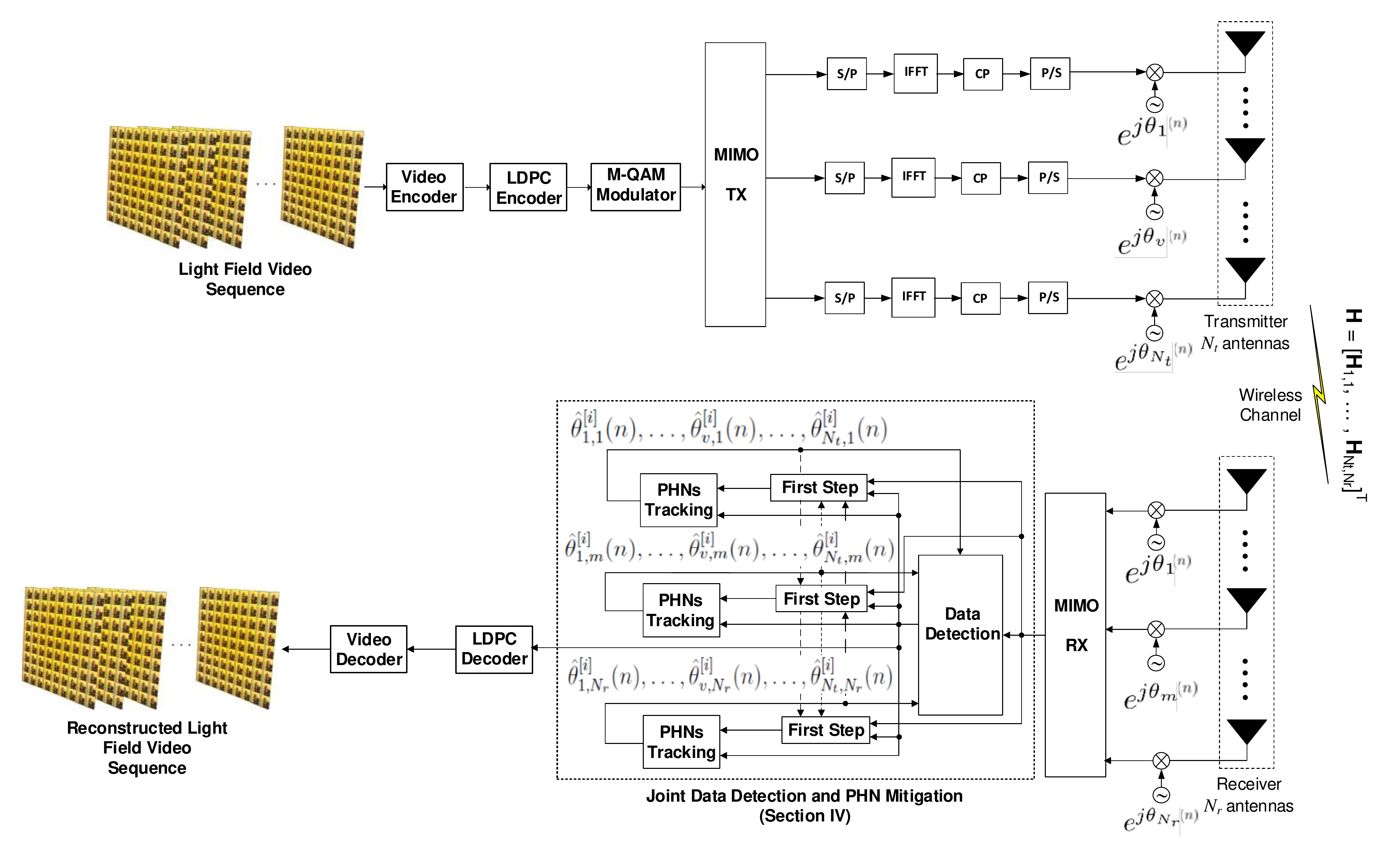}
	\end{center}
	\caption{Block diagram of a MIMO-OFDM system with proposed detector based on an EKF algorithm and data detection for light field video transmission.}
	\label{fig:Systemmodel}
\end{figure}

Fig. \ref{fig:Systemmodel} shows the diagram of a MIMO-OFDM system for LF video transmission equipped with $N_t$ transmit and $N_r$ receive antennas. The following subsections describe major components of the proposed system. 
\subsection{LF Video Encoding and Decoding} \label{sec:LFvideENandDE}
\begin{figure}[h]
	\begin{center}
		\includegraphics[width=0.7\textwidth]{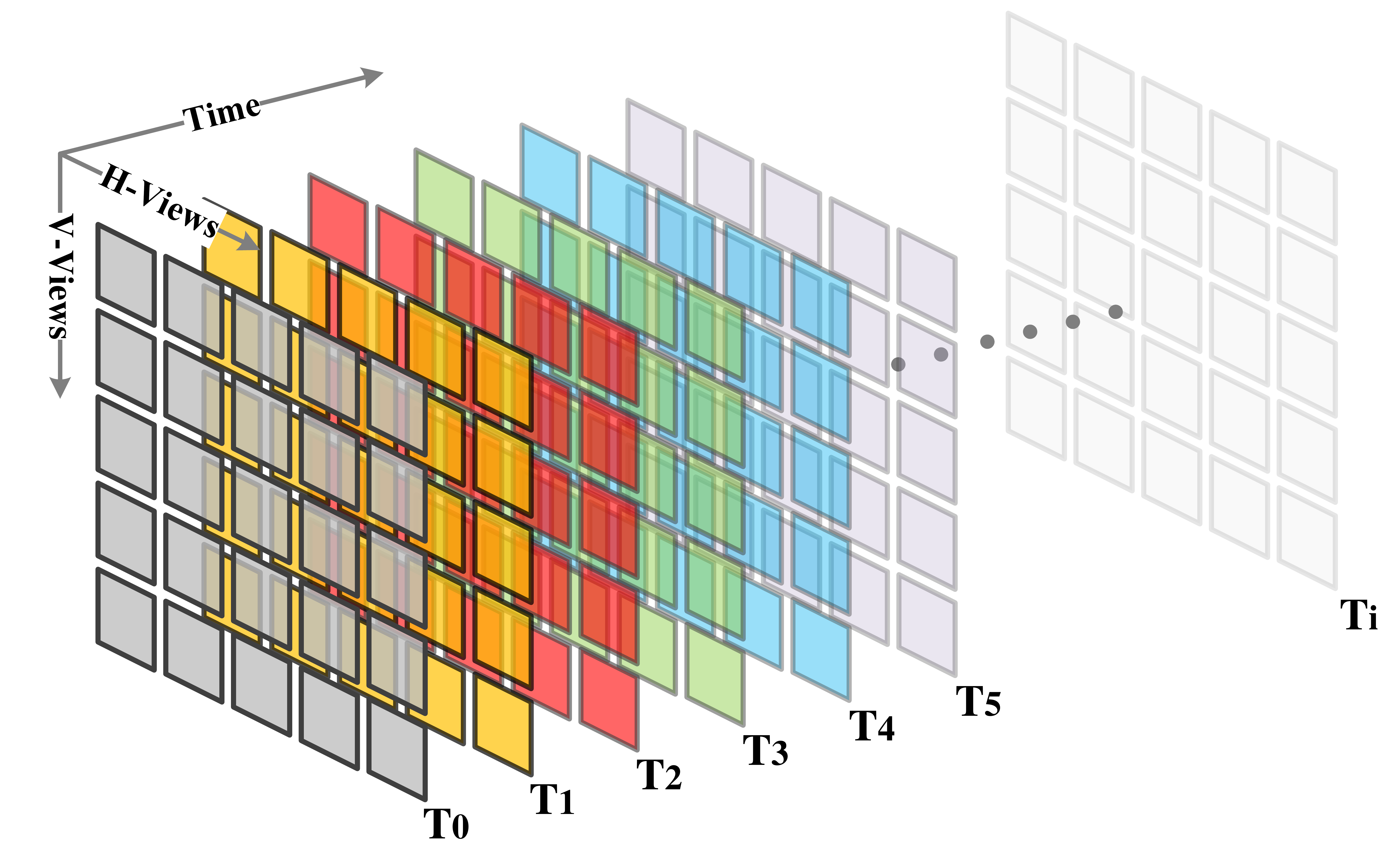}
	\end{center}
	\caption{An LF video sequence represented by a 3-D image matrix.}
	\label{fig:LFV}
\end{figure}

Compared to the conventional 3-D formats, the main problem of LF video transmission in wireless systems is the huge amount of video data, since LF video exhibits a more complex structure and consists of densely distributed 2-D image arrays. Therefore, state-of-the-art video compression techniques such as H.264/AVC and H.264/MVC are required to compress LF video.
As shown in Fig. \ref{fig:LFV}, an LF video can be represented as a 3-D image matrix, with three dimensions in horizontal views (H-Views), vertical views (V-Views), and time.  

\begin{figure}[h]
	\begin{center}
		\includegraphics[width=0.9\textwidth]{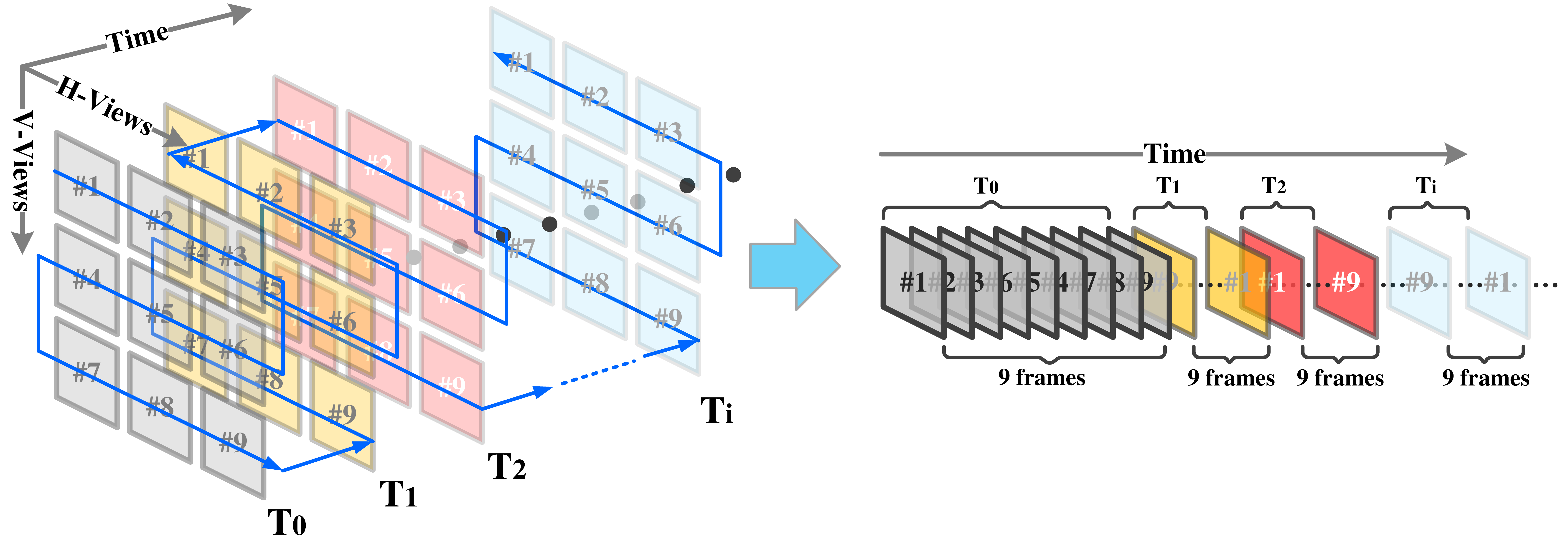}
	\end{center}
	\caption{Transposed picture ordering for LF video coding.}
	\label{fig:TPC}
\end{figure}

Fig. \ref{fig:TPC} shows an encoding example of an LF video with $3 \times 3$ views. LF video encoding can be achieved by transposing the 3-D image matrix into a single view video sequence, which is then encoded with H.264/AVC encoder. More specifically, as shown in Fig. \ref{fig:TPC},  
the first LF frame is transposed from the top-left to bottom-right, then the second LF frame is transposed from bottom-right to top-left, and so on. 

\subsection{MIMO-OFDM System Model} \label{sec:LFvideENandDE}
Here, the following set of assumptions are adopted:
\begin{description}
	\item[A1.] The channels are modeled as a slow fading frequency-selective channel, i.e., the channels are assumed to be quasi-static Rayleigh fading, which are constant over group of pictures (GOP) data packet, and known over the OFDM packet duration and change from packet to packet following a complex Gaussian distribution.
	\item[A2.] The time-varying PHN change from symbol to symbol and is modeled as a Wiener process, i.e., $\theta_n=\theta_{n-1}+\delta_n$, $\forall$ $n$, where $\theta_n$ is the PHN at the $n$th instant, $\delta_n \sim \mathcal{N}(0,\sigma^2_\delta)$ is PHN innovation, and $ \sigma^2_\delta $ is the variance of the innovation process \cite{{Demir:2000May},{Lin:2006apr}}.
	\item[A3.] In order to ensure generality, each transmit and receive antenna are assumed to be equipped with an independent oscillator as depicted in Fig. 1. This ensures that the system model is in line with previous work in \cite{{Hadaschik:2005Mar},{Hani:2012sept}} and is also applicable to various MIMO scenarios, e.g., line-of-sight (LoS) MIMO and SDMA MIMO systems.
	\item[A4.] Channels and CFOs are assumed to be estimated and compensated during the transmission of training sequences. 
	\item[A5.] The timing mismatch is assumed to not exceed the length of the cyclic prefix and result in ICI. Hence, it is not considered here.
\end{description}
Note that assumptions A1, A2, A3, A4 and A5 are in line with previous studies and channel, and PHN estimation algorithms in \cite{{Zhang:2009Sep},{Huang:2010Jul},{Yao:2012Dec},{Rabiei:2011May},{Lin:2006apr},{Lin:2006sept},{Lin:2007May},{Rabiei:20010Nov},{Carvajal:2013Jan},{DaweiWang:2013}}. Assumption A1 is adopted in the IEEE $802.11$ac/ad standards in \cite{{Wu:2002Dec},{Munier:2008Jul},{Lin:2006sept},{Petrovic:2007Aug},{Rabiei:20010Nov},{Paul:2008FQ},{Zhu:2011May},{DaweiWang:2013}}. Assumption A2 is also reasonable in many practical scenarios to
describe the behavior of practical oscillators \cite{{Demir:2000May},{Lin:2006sept}}. In addition, assumption A4 is reasonable since the channels and CFOs are assumed to be estimated during the training process using algorithms proposed in \cite{{Minn:2005Dec},{Minn:2006Oct}}.

The time-invariant composite channel impulse response (CIR) between any pair of transmit antenna $v$ and receive antenna $m$ is modeled as $h_{v,m}(\tau)=\sum_{l=0}^{L-1}h_{v,m}(l)\delta(\tau-lT_{sp})$, where $h_{v,m}(l)$ is the channel gain for the \emph{l}th tap, and $\delta(x)$ is the unit impulse function. \emph{L} denotes the number of channel taps, and $T_{sp}=1/B$, is the sampling time. For brevity, it is assumed that the channel order \emph{L} is the same for any pair of antennas \cite{Chi:2011Jul}. The frequency-domain channel coefficient matrix is $\mathbf{H}_{v,m}= \text{diag}\{H_{v,m}(0),H_{v,m}(1),\ldots,H_{v,m}(N-1)\}$, where $H_{v,m}{(n)}=\sum_{d=0}^{L-1}h_{v,m}(d)e^{-(j2 \pi nd/N)}$ is the channel frequency response on the \emph{n}th subcarrier and $N$ is the number of subcarriers.
The baseband transmitted OFDM signal can be obtained as follows. A set of modulated video data $\mathbf{X}=[X(0),X(1),\cdots,X(N-1)]^T\in \mathcal{C}^{N \times 1}$ of LF video sequence after phase-shift keying (PSK) or quadrature amplitude modulation (QAM) is normalized by IDFT, at the $v$th antenna, as
\begin{eqnarray}
x(n)=\frac{1}{\sqrt{N}}\sum\limits^{N-1}_{k=0}X(k)e^{\frac{j2 \pi kn}{N}} &&n=0,1,\ldots,N-1,
\end{eqnarray}
where $k$ denotes the $k$th symbol of the OFDM symbol and $N$ is the number of subcarriers. The received baseband signal of an OFDM symbol at the $m$th antenna in time domain, after removing the cyclic prefix (CP), is given by
\begin{align} \label{eq:recsigkant}
y_m(n)=\frac{1}{\sqrt{N}}\sum_{v=1}^{N_t}e^{j\theta_{v,m}(n)}\sum\limits^{N-1}_{k=0}H_{v,m}(k) X(k) e^{j2 \pi kn/N}+w_m(n),  \ \ \ \ m=1,\dots,N_r \nonumber\\
\end{align}
where $\{\theta_{v,m}(n)=\theta_v(n)+\theta_m(n)\}^{N-1}_{n=0}$ is the PHN process between $v$th and $m$th antennas, $\{H_{v,m}^{(k)}\}^{N-1}_{k=0}$ is the channel frequency response between $v$th and $m$th antennas for subcarriers $0$ to $N-1$, and $\{w_m(n)\}^{N-1}_{n=0}$ is the complex additive white Gaussian noise (AWGN) with variance $\sigma_w^2$ at the $m$th antenna.
\section{Problem Formulation}

The received data signal at the $m \rm th$ antenna in \eqref{eq:recsigkant} can be written as \cite{Petrovic:ICC2005May,Petrovic:2007Aug}
\begin{align} \label{eq:ymmatrixform}
\mathbf{y}_m=\sum_{v=1}^{N_t} \mathbf{P}_{v,m}(\mathbf{x} \odot \mathbf{h}_{v,m})+\mathbf{w}_m, \ \ \ m=1,\dots,N_r
\end{align}
where
\begin{itemize}
	\item $\mathbf{x} \triangleq \text{IDFT}\{\mathbf{X}\}$ is the data vector in the time domain,
	\item $\mathbf{P}_{v,m} \triangleq \text{diag}([e^{j \theta_{v,m}(0)}, e^{j \theta_{v,m}(1)},\ldots,e^{j \theta_{v,m}(N-1)}]^T)$ is the $N \times N$ PHN matrix between any pair of antennas $v$ and $m$,
	\item $\mathbf{h}_{v,m} \triangleq [h_{v,m}(0),h_{v,m}(1),\ldots,h_{v,m}(L-1)]^T$ is the channel impluse response (CIR) with the channel order, $L$, and the channel gains, $h_{v,m}(l)$ are modeled as complex Gaussian zero-mean random variables,
	\item $\mathbf{w}_m=[w_m(0),w_m(1),\ldots,w_m(N-1)]^T$ is the zero-mean complex additive white Gaussian noise (AWGN) vector at the $m$th receive antenna, i.e., $w_m(n) \sim \mathcal{N}(0,\sigma^2_w)$ and $\sigma^2_w$ is the AWGN variance.
\end{itemize}

At the receiver, after DFT, the received signal in \eqref{eq:ymmatrixform} is given by \cite{Petrovic:ICC2005May,Petrovic:2007Aug}
\begin{align} \label{eq:finalrecsigitrJ}
\mathbf{Y}_m&=\sum_{v=1}^{N_t} \mathbf{J}_{v,m} \odot (\mathbf{H}_{v,m} \mathbf{X})+\boldsymbol{\eta}_m, 
=\mathbf{J}_{v,m} \odot (\mathbf{H}_{v,m} \mathbf{X})+\mathbf{Z}_m+\boldsymbol{\eta}_m,
\end{align}
where
\begin{itemize}
	\item $\mathbf{Z}_m$ is the interference signal from the transmit antennas other than the $v$th transmit antenna and is given by
	\begin{align} \label{eq:intersig}
	\mathbf{Z}_m = \sum_{\substack{\ell=1\\\ell \neq v}}^{N_t} \mathbf{J}_{\ell,m} \odot (\mathbf{H}_{\ell,m} \mathbf{X}),
	\end{align}
	\item In \eqref{eq:finalrecsigitrJ} and \eqref{eq:intersig}, $\mathbf{J}_{a,m} \triangleq [J_{a,m}(0),J_{a,m}(1),\ldots,J_{a,m}(N-1)]^T$ for $a \in \{v,\ell\}$ is the DFT coefficients of the PHN vector, where its $k$th coefficient is given as
	\begin{eqnarray} \label{eq:eqJam}
	J_{a,m}(k)=\frac{1}{N}\sum\limits^{N-1}_{n=0}e^{j\theta_{a,m}(n)}e^{\frac{-j2 \pi kn}{N}} &&k=0,1,\ldots,N-1,
	\end{eqnarray}
	\item $\boldsymbol{\eta}_m \triangleq \text{DFT}\{\mathbf{w}_m\}$.
\end{itemize}
In  \eqref{eq:finalrecsigitrJ}, the received signal corresponding to the \emph{k}th subcarrier, $Y_m(k)$, is given by \cite{Petrovic:ICC2005May,Petrovic:2007Aug}
\begin{align} \label{eq:ymkafterFFT}
Y_m(k)&=\underbrace{J_{v,m}(0)}_{\text{CPE}}\underbrace{H_{v,m}(k)X(k)}_{\text{desired subcarrier signal}}+\sum_{\substack{\vartheta=0\\\vartheta \neq k}}^{N-1} \underbrace{J_{v,m}(k-\vartheta)H_{v,m}(\vartheta)X(\vartheta)}_{\text{ICI}} \nonumber\\
&+Z_m(k)+\eta_m(k),
\end{align}
where $Z_m(k)$ is the $k$th symbol of the vector, $\mathbf{Z}_m$ in \eqref{eq:intersig}.

It is clearly seen in \eqref{eq:ymkafterFFT} that the desired subcarrier signal, $H_{v,m}(k) X(k)$, is affected by the interference signals produce from the CPE, $J_{v,m}(0)$, the ICI, $\sum_{\substack{\vartheta=0\\\vartheta \neq k}}^{N-1} J_{v,m}(k-\vartheta)H_{v,m}(\vartheta)X(\vartheta)$, and the interference signals from other antennas, $Z_m(k)$ in \eqref{eq:intersig}. 
As will be shown in (\ref{eq:eqDT}) and (\ref{eq:eqBk}), in the next section, the accuracy of estimating $\text{ICI}$, $\text{CPE}$, and $Z_m(k)$ in \eqref{eq:ymkafterFFT} 
dramatically impacts the overall PSNR performance of a MIMO system. Therefore, in this paper, we focus on estimating the PHN parameters for all transmit antennas in the time domain and the desired data subcarrier, $X(k)$, in the frequency domain. Then, we will determine the interfering items, CPE, ICI, and the interference signal from other antennas, $Z_m(k)$. By cancelling these interference items, the data detection will be improved and the video distortion will be consequently reduced. Thus, we propose an iterative algorithm based on the EKF to estimate the data in the presence of multiple PHN parameters.        

\subsection{LF Video Transmission in MIMO-OFDM Systems}

To describe the instantaneous rate of the source, we model the rate-distortion (RD) video curve with a parameterized function. Let $D_T$ denotes the RD function and the distortion MSE is approximated by \cite{{Stuhlmuller:2000},{DaweiWang:2013},{DaweiWang:2015}}
\begin{align} \label{eq:eqDT}
D_T=\frac{b}{\frac{T_s}{T_0}\sum_{v=1}^{N_t}\sum_{k=0}^{N-1}\hat{R}_{v}{(k)}+z}+a,
\end{align}
where
\begin{itemize}
	\item $a$, $b$ and $z$ are constants and depend on the light field video content. The RD function is relatively flat and the $b$ parameter is relatively small when the video motion is relatively slow. However, the RD function is normally steep and the $b$ parameter is relatively large for fast video motion \cite{{Stuhlmuller:2000},{DaweiWang:2013},{DaweiWang:2015}},
	\item $T_s$ is the duration of one GOP,
	\item $T_0=T+T_{cp}$, $T$ is the data duration, $T_{cp}$ is the duration of the cyclic prefix,
	\item $\hat{R}_{v}{(k)}$ is the estimated instantaneous rate of the source at the $v$th antenna and that is given by \cite{{DaweiWang:2013},{DaweiWang:2015}} 
	\begin{align} \label{eq:eqBk}
	\hat{R}_{v}{(k)}=r_c \ \log_2\left[1+\frac{|\hat{X}(k)|^2\sum_{m=1}^{N_r}|H_{v,m}{(k)}|^2}{\sum_{m=1}^{N_r}\big|Y_m(k)-\hat{S}_m(k)-\hat{S}_m^{\text{int}}(k)\big|^2}\right],
	\end{align}
	\item $\hat{X}(k)$ is the estimated desired data subcarrier from (cf. \ref{eq:eqLShestforX_1}), in the next section,
	\item $\hat{S}_m(k)$ is the estimate of the $k$th symbol of the vector, $\hat{\mathbf{S}}_m(k) \triangleq \hat{\mathbf{J}}_{v,m} \odot (\mathbf{H}_{v,m} \hat{\mathbf{X}})$,
	\item $\hat{\mathbf{J}}_{v,m} \triangleq [\hat{J}_{v,m}(0),\hat{J}_{v,m}(1),\ldots,\hat{J}_{v,m}(N-1)]^T$ is the estimated DFT coefficients of the PHN vector, where its $k$th coefficient is given by
	\begin{eqnarray} \label{eq:eqJam}
	\hat{J}_{v,m}(k)=\frac{1}{N}\sum\limits^{N-1}_{n=0}e^{j\hat{\theta}_{v,m}(n)}e^{\frac{-j2 \pi kn}{N}} &&k=0,1,\ldots,N-1,
	\end{eqnarray}
	\item $\hat{\theta}_{v,m}(n)$ is the estimate $n$th PHN symbol from (cf. \ref{eq:eq16}), in the next section,
	\item $r_c$ is the channel code rate for protecting the video data against the noise in the channel.
\end{itemize}

%

It is clearly shown from \eqref{eq:eqDT} and \eqref{eq:eqBk} that the video distortion, $D_T$, is only determined by the estimation accuracy of a data vector, $\hat{\mathbf{X}}$, and PHN parameters, $\hat{\boldsymbol{\theta}}_{v,m}$, between any pair of transmit antenna $v$ and receive antenna
$m$. As shown in Section \ref{sec:Complexityanalysis} and will be shown in Section \ref{sec:Simresultsdis}, the estimation accuracy of $\hat{\mathbf{X}}$ and $\hat{\boldsymbol{\theta}}_{v,m}$ dramatically impacts the overall PSNR performance. 
In Section \ref{sec:Simresultsdis}, we will provide simulation results to show that there are additional factors related to estimation accuracy of $\hat{\mathbf{X}}$ and $\hat{\boldsymbol{\theta}}_{v,m}$ that directly affects the performance of MIMO-OFDM systems for light field video transmission and have to be considered in the design of multimedia MIMO-OFDM communications systems. 
\section{Joint data detection and PHN mitigation} \label{sec:Itrest}
In this section, an iterative algorithm based on the EKF that jointly detects data and estimates multiple PHN parameters is derived. The received signal in \eqref{eq:recsigkant}, $\mathbf{y}_m=[y_m(0),y_m(1),\ldots,y_m(N-1)]^T$, may be written in matrix form as \cite{Lin:2006sept,Lin:2007May}
\begin{subequations}\label{eq:recsigkantab}
	\begin{align}
	\mathbf{y}_m&=\sum_{v=1}^{N_t} \bar{\mathbf{P}}_{v,m}\mathbf{F}^H\bar{\mathbf{H}}_{v,m}\mathbf{d}_v+\mathbf{w}_m, \label{eq:recsigkanta} \\
	&=\sum_{v=1}^{N_t} \mathbf{P}_{v,m}\mathbf{F}^H\mathbf{H}_{v,m}\mathbf{d}_v+\mathbf{w}_m, \label{eq:recsigkantb} \ \ \ \ m=1,\dots,N_r
	\end{align}
\end{subequations}
where
\begin{itemize}
	\item $H_{v,m} (n) \triangleq e^{j\bar{\theta}_{v,m} (0)} \bar{H}_{v,m} (n)$, and $\theta_{v,m} (n) \triangleq \bar{\theta}_{v,m} (n) - \bar{\theta}_{v,m} (0)$. This equivalent system model in \eqref{eq:recsigkantb} helps to distinguish between the phase disturbance caused by PHN and the channel phase for the first sample,
	\item $\mathbf{F}_L$ is the $N \times L$ DFT matrix, with $\mathbf{F}_L \triangleq \mathbf{F}(1:N,1:L)$,
	\item $\mathbf{H}_{v,m} \triangleq \text{diag}(\mathbf{F}_L\mathbf{h}_{v,m})$ is the channel frequency response,
	\item Following \cite{Chi:2011Jul}, $\mathbf{d}_v \triangleq \boldsymbol{\Lambda}_v \mathbf{X}$, $\boldsymbol{\Lambda}_v = \text{diag}(1,e^{{j2 \pi (L+1) v}/N},\ldots,e^{{j2 \pi (L+1)(N-1) v }/N})$, and $\mathbf{X}=[X(0),X(1),\ldots,X(N - 1)]^T$ is the transmitted data vector. As explained in \cite{Chi:2011Jul}, the transform operator $\boldsymbol{\Lambda}_v$ can be viewed as a frequency modulation and could be used to achieve the orthogonality between the received signals at the receiver,
	\item  $\mathbf{F}$ is an $N \times N$ DFT matrix, with  $ [\mathbf{F}]_{\emph{l},\emph{p}} \triangleq (1/ \sqrt{N}) e^{-j(2 \pi pl/N)}$ for $p,l=0,1,\cdots,N-1$,
	\item $\mathbf{w}_m=[w_m(0),w_m(1),\ldots,w_m(N-1)]^T$ is the zero-mean complex additive white Gaussian noise (AWGN) vector at the $m$th receive antenna, i.e., $w_m(n) \sim \mathcal{N}(0,\sigma^2_w)$ and $\sigma^2_w$ is the AWGN variance.
\end{itemize}
Eq. \eqref{eq:recsigkantb} can be written as follows
\begin{equation} \label{eq:eqrecsigkant}
\mathbf{y}_m =\underbrace{\big[\boldsymbol{\Psi}_1,\boldsymbol{\Psi}_2,\ldots,\boldsymbol{\Psi}_{N_t}\big]}_{\boldsymbol{\Psi}} \ \mathbf{X} +\mathbf{w}_m,
\end{equation}
where $\boldsymbol{\Psi} \triangleq \big[\boldsymbol{\Psi}_1,\boldsymbol{\Psi}_2,\ldots,\boldsymbol{\Psi}_{N_t}\big],$
is an $N \times NN_t$ matrix, and $\boldsymbol{\Psi}_v \triangleq \mathbf{P}_{v,m}\textbf{F}^H\mathbf{H}_{v,m}\boldsymbol{\Lambda}_v$, for $v=1,\ldots,N_t$, is an $N \times N$ matrix.

The combined received signal at all $N_r$ antennas can be written as
\begin{equation} \label{eq:eqrecsigNrant}
\mathbf{y} =\boldsymbol{\Psi} \ \mathbf{X} +\mathbf{w}
\end{equation}
where $\mathbf{y} \triangleq \big[\mathbf{y}_1^T,\mathbf{y}_2^T,\ldots,\mathbf{y}_{N_r}^T\big]^T$ is an $N \times N_r$ matrix, and $\mathbf{w} \triangleq [\mathbf{w}_1^T,\ldots,\mathbf{w}_{N_r}^T]^T$ is an $N \times N_r$ AWGN matrix.

If the PHN vector, $\boldsymbol{\theta}$, is known, the data vector in \eqref{eq:eqrecsigNrant} can be estimated via an LS estimator as, $\hat{\mathbf{X}}=(\boldsymbol{\Psi}^H\boldsymbol{\Psi})^{-1}\boldsymbol{\Psi}^H\mathbf{y}$. Then, the estimate of $\boldsymbol{\theta}$ is obtained via  
\begin{align} \label{eq:eqseachPHNsCFOs}
\hat{\boldsymbol{\theta}} = \arg \underset{\boldsymbol{\theta}}{\max} \ \mathbf{y}^H\boldsymbol{\Psi}(\boldsymbol{\Psi}^H\boldsymbol{\Psi})^{-1}\boldsymbol{\Psi}^H\mathbf{y},
\end{align}
where $\boldsymbol{\theta} \triangleq [\boldsymbol{\theta}_1^T,\ldots,\boldsymbol{\theta}_{N_r}^T]^T$ is a $N_tN \times N_r$ matrix PHN, $\boldsymbol{\theta}_m = [\boldsymbol{\theta}_{1,m}^T,\ldots,\boldsymbol{\theta}_{N_t,m}^T]^T$, for $m=1,\ldots,N_r$, is an $N_tN \times 1$ vector, and $\boldsymbol{\theta}_{v,m} =[\theta_{v,m},\dots,\theta_{v,m}]^T$ is an $N \times 1$ vector.

The maximization in \eqref{eq:eqseachPHNsCFOs} needs to be carried out using a multidimensional exhaustive search over the set of possible PHNs. This operation would require an exhaustive search over the multi-dimensional space spanned by $\boldsymbol{\theta}$, which is prohibitively complex in practical systems. Thus, in order to reduce the computational complexity associated with obtaining these estimates, we propose an iterative scheme based on the EKF estimator. In the first step, the signal of each transmit antenna, i.e., $\hat{\mathbf{y}}_v$ ($v=1,2,\ldots,N_t$), is first extracted from the received vector $\mathbf{y}_m$. In the second step, each received signal of $v$th antenna, $\hat{\mathbf{y}}_v$, is then exploited to jointly estimate $\boldsymbol{\theta}_{v,m}$ and $\mathbf{X}$.

\textbf{First-step} \\
Compute
\begin{align} \label{eq:SAGE_Estep}
\hat{\mathbf{y}}_v^{[i]} = \mathbf{y}_m-\sum_{\substack{\ell=1\\\ell \neq v}}^{N_t} \hat{\mathbf{P}}_{\ell,m}^{[i]}\mathbf{F}^H\hat{\mathbf{H}}_{\ell,m}\mathbf{d}_\ell^{[i]}.
\end{align}
Substituting \eqref{eq:recsigkantb} into \eqref{eq:SAGE_Estep} yields
\begin{align} \label{eq:eqfinalcompdata}
\hat{\mathbf{y}}_v^{[i]} & = \mathbf{P}_{v,m}\mathbf{F}^H\mathbf{H}_{v,m}\mathbf{d}_v \nonumber\\
&+ \underbrace{\sum_{\substack{\ell=1\\\ell \neq v}}^{N_t} \bigg[\mathbf{P}_{\ell,m}\mathbf{F}^H\mathbf{H}_{\ell,m}\mathbf{d}_v-
	\hat{\mathbf{P}}_{\ell,m}^{[i]}\mathbf{F}^H\mathbf{H}_{\ell,m}\mathbf{d}_\ell^{[i]}\bigg]+\mathbf{w}}_{\textbf{overall noise} \ \triangleq \ \boldsymbol{\alpha}^{[i]}_v},
\end{align}
where the overall noise vector is defined by $\boldsymbol{\alpha}^{[i]}_v \triangleq [\alpha^{[i]}_v(0),\ldots,\alpha^{[i]}_v(N-1)]^T$.
Following \cite{Pun:2007Feb}, the overall noise vector, $\boldsymbol{\alpha}^{[i]}_v$, is a disturbance term that accounts
for thermal noise and residual interference from the antennas after the $i$th iteration, where the interference from the antennas is linearly related to the data symbols of all interfering antennas. Then, assuming that the latter is independent and identically distributed with zero-mean, it follows from the central limit theorem that the entries of $\boldsymbol{\alpha}^{[i]}_v$ are nearly Gaussian distributed with zero-mean and some variance $\sigma_w^2$. Under the above assumption, it turns out that the minimization problem in \eqref{eq:SAGE_Mstep} is equivalent to the ML estimation of $\boldsymbol{\theta}_{v,m}$ and $\mathbf{X}$ starting from the observations $\hat{\mathbf{y}}_v^{[i]}$, for $v=1,\dots,N_t$.

\textbf{Second-step} \\
Compute
\begin{align} \label{eq:SAGE_Mstep}
\hat{\boldsymbol{\lambda}}_v^{[i]} = \arg \underset{\boldsymbol{\lambda}_v}{\min} \bigg\|\hat{\mathbf{y}}_v^{[i]}-\mathbf{P}_{v,m}\mathbf{F}^H\mathbf{H}_{v,m}\mathbf{d}_v \bigg\|^2.
\end{align}
%
%


At each stage of the iterative algorithm, only one parameter is updated while the others are kept constant at their most updated values. This makes the iterative algorithm suitable for multi-dimensional ML estimation problems, where the likelihood function must be optimized over several parameters \footnote{Convergence is addressed in Section IV.A}. In the following, we apply the iterative algorithm based on the EKF estimator to solve the optimization problem in \eqref{eq:SAGE_Mstep}.

The signal, $\hat{y}_v^{[i]} (n)$ in \eqref{eq:SAGE_Estep} is used to estimate the PHN vector as follows. The state and observation equations at time \emph{n} are given by
\begin{align} \label{eq:eq11}
&\theta_{v,m}(n) = \theta_{v,m}({n-1})+\delta_{v,m}(n),\\
\label{eq:eq7}
&y_v (n)=g_v (n)+\alpha_v (n)=e^{j\theta_{v,m} (n)}s_v^{[i]} (n)+\alpha_v (n),
\end{align}
where $s_v^{[i]}(n)$ is the $n$th symbol of the vector, $\mathbf{s}_v^{[i]} \triangleq \mathbf{F}^H\mathbf{H}_{v,m}\hat{\mathbf{d}}_v^{[i]}$, $\hat{\mathbf{d}}_v^{[i]} \triangleq \boldsymbol{\Lambda}_v \hat{\mathbf{X}}^{[i]}$, $\hat{\mathbf{X}}^{[i]}$ is estimated by an MMSE linear receiver as
\begin{align} \label{eq:eqLShestforX}
\hat{\mathbf{X}}^{[i]}=(\boldsymbol{\Psi}^{[i]H}\boldsymbol{\Psi}^{[i]}+\sigma_w^2\mathbf{I}_{NN_t \times NN_t})^{-1}\boldsymbol{\Psi}^{[i]H}\mathbf{y},
\end{align}
$\boldsymbol{\Psi}^{[i]}=\big[\boldsymbol{\Psi}_1^{[i]},\boldsymbol{\Psi}_2^{[i]},\ldots,\boldsymbol{\Psi}_{N_t}^{[i]}\big]$, $\boldsymbol{\Psi}_v^{[i]} \triangleq \mathbf{P}_{v,m}^{[i]}\textbf{F}^H\mathbf{H}_{v,m}\boldsymbol{\Lambda}_v$ for $v=1,\ldots,N_t$, and $\mathbf{P}_{v,m}^{[i]} \triangleq \text{diag}([e^{j \theta_{v,m}^{[i]}(0)},$ $ e^{j \theta_{v,m}^{[i]}(1)},\ldots,e^{j \theta_{v,m}^{[i]}(N-1)}]^T)$ is the PHN matrix between any pair of antennas $v$ and $m$.

Since the observation equation in (\ref{eq:eq7}) is a non-linear function of the unknown state vector $\boldsymbol{\theta}_{v,m}$, the EKF is used instead of a simple Kalman filter. The EKF uses Taylor series expansion to linearize the non-linear observation equation in (\ref{eq:eq7}) about the current estimates \cite{Kay:1993}. Thus, the Jacobian of $g_v (n)$ is evaluated by computing the first order partial derivative of $g_v (n)$ with respect to $\theta_{v,m} (n)$ as
\begin{align} \label{eq:eq12}
\dot{g}_v (n)=\frac{\partial g_v (\theta_{v,m} (n))}{\partial \theta_{v,m} (n)} \bigg|_{\theta_{v,m} (n)=\hat{\theta}_{v,m} ({n|n-1})}
&= j g (\hat{\theta}_{v,m} ({n|n-1})) 
= j e^{j\hat{\theta}^{[i]}_{v,m} ({n|n-1})}\hat{s}_v^{[i]} (n),
\end{align}
where $\dot{g}$ denotes the Jacobian of $g$ evaluated at $\theta_{v,m} (n)$. The first and second moments of the state vector at the $(i+1)$th iteration denoted by $\hat{\theta}_{v,m}^{[i+1]} ({n|n-1})$ and $M^{[i+1]} ({n|n-1})$, respectively, are given by
\begin{align} \label{eq:eq13}
\hat{\theta}^{[i+1]}_{v,m} ({n|n-1})&=\hat{\theta}^{[i+1]}_{v,m} ({n-1|n-1}),
\\
\label{eq:eq14}
M^{[i+1]} ({n|n-1})&=M^{[i+1]} ({n-1|n-1})+\sigma^2_{\delta_{v,m}}.
\end{align}
Given the observation $y_v (n)$, the Kalman gain $K(n)$, posteriori state estimate $\hat{\theta}^{[i+1]}_{v,m} ({n|n})$, and the filtering error covariance, $M^{[i+1]} ({n|n})$ are given by
\begin{align} \label{eq:eq15}
K (n)&=M^{[i+1]} ({n|n-1})\dot{g}^\ast(\theta_{v,m} ({n|n-1})) \nonumber \\
& \times \Big(\dot{g}(\theta_{v,m} ({n|n-1})) \times M^{[i+1]} ({n-1|n-1})
\times \dot{g}^\ast(\theta_{v,m} ({n|n-1}))+ \sigma^2_w\Big)^{-1}, \\
\label{eq:eq16}
\hat{\theta}^{[i+1]}_{v,m} ({n|n}) &=\hat{\theta}^{[i+1]}_{v,m} ({n|n-1})
+\Re\big\{K_n\big(u_v (n)-e^{j\hat{\theta}^{[i+1]}_{v,m} ({n|n-1})}\hat{s}_v^{[i]} (n)\big)\big\},
\\
\label{eq:eq17}
M^{[i+1]} ({n|n})&=\Re\big\{M^{[i+1]} ({n|n-1})-K (n)\dot{g}(\theta_{v,m} ({n|n-1})) \times
M^{[i+1]} ({n|n-1})\big\},
\end{align}
respectively. Before starting the EKF recursion (\ref{eq:eq12})-(\ref{eq:eq17}), $\hat{\theta}^{[1]}_{v,m} ({1|0})$ and $M^{[1]} ({1|0})$ are initialized by $\hat{\theta}^{[1]}_{v,m} ({1|0})=0$ and $M^{[1]} ({1|0})=\sigma^2_{\delta_{v,m}}$.
Using \eqref{eq:SAGE_Estep}-\eqref{eq:eq17}, we can estimate the PHN parameters from all antennas, i.e., $\hat{\boldsymbol{\theta}}^{[i+1]} \triangleq [\hat{\boldsymbol{\theta}}_1^{[i+1]T},\ldots,\hat{\boldsymbol{\theta}}_{N_r}^{[i+1]T}]$, $\hat{\boldsymbol{\theta}}_m^{[i+1]} = [\hat{\boldsymbol{\theta}}_{1,m}^{[i+1]T},\ldots,\hat{\boldsymbol{\theta}}_{N_t,m}^{[i+1]T}]$ for $m=1,\ldots,N_r$, and $\hat{\boldsymbol{\theta}}_{v,m}^{[i+1]} =[\hat{\theta}_{v,m}^{[i+1]},\dots,\hat{\theta}_{v,m}^{[i+1]}]^T$. 
Then, the data vector can be updated as
\begin{align} \label{eq:eqLShestforX_1}
\hat{\mathbf{X}}^{[i+1]}=(\boldsymbol{\Psi}^{[i+1]H}\boldsymbol{\Psi}^{[i+1]}+\sigma_w^2\mathbf{I}_{NN_t \times NN_t})^{-1}\boldsymbol{\Psi}^{[i+1]H}\mathbf{y},
\end{align}
where $\boldsymbol{\Psi}^{[i+1]}=\big[\boldsymbol{\Psi}_1^{[i+1]},\boldsymbol{\Psi}_2^{[i+1]},\ldots,\boldsymbol{\Psi}_{N_t}^{[i+1]}\big]$, $\boldsymbol{\Psi}_v^{[i+1]} \triangleq \mathbf{P}_{v,m}^{[i+1]}\textbf{F}^H\mathbf{H}_{v,m}\boldsymbol{\Lambda}_v$ for $v=1,\ldots,N_t$, and $\mathbf{P}_{v,m}^{[i+1]} \triangleq \text{diag}([e^{j \theta_{v,m}^{[i+1]}(0)},$ $ e^{j \theta_{v,m}^{[i+1]}(1)},\ldots,e^{j \theta_{v,m}^{[i+1]}(N-1)}]^T)$.
Using \eqref{eq:SAGE_Estep} - \eqref{eq:eqLShestforX_1} and reapplying the above algorithm, for $v=\{1,\ldots,N_t\}$, the iterations of the proposed algorithm stop when the difference between LLFs of two iterations is smaller than a threshold $\zeta$, i.e.
\begin{align} \label{eq:eq25}
\Bigg |\bigg\| \mathbf{y}_m-\sum_{v=1}^{N_t} \hat{\mathbf{P}}_{v,m}^{[i+1]}\boldsymbol{\Gamma}_v\hat{\mathbf{X}}^{[i+1]} \bigg\| ^2
- \bigg\| \mathbf{y}_m-\sum_{v=1}^{N_t} \hat{\mathbf{P}}_{v,m}^{[i]}\boldsymbol{\Gamma}_v\hat{\mathbf{X}}^{[i]} \bigg\| ^2 \Bigg | \leq \zeta.
\end{align}
where $\boldsymbol{\Gamma}_v \triangleq \textbf{F}^H\mathbf{H}_{v,m}\boldsymbol{\Lambda}_v$.
\subsection{Initialization and Convergence} \label{sec:InitialConver}
Let $\hat{\mathbf{X}}^{[0]}$ denote the initial estimate of the transmitted data vector. Appropriate initialization of $\hat{\mathbf{X}}^{[0]}$ results in the proposed iterative detector to converge quickly. In our algorithm, the initial data estimate is obtained using
$\hat{\mathbf{X}}^{[0]}=(\boldsymbol{\Psi}^{[j-1]H}\boldsymbol{\Psi}^{[j-1]})^{-1}\boldsymbol{\Psi}^{[j-1]H}\mathbf{y}$, where $\hat{\mathbf{P}}_{v,m}^{[j-1]}$ in \eqref{eq:eqrecsigkant} is the PHN matrix estimate obtained from the previous OFDM symbol. Simulation results in Section~\ref{sec:Simresultsdis} indicate that at SNR$=25$ dB the proposed detector, on average, converges after $2$ iterations.

\emph{Remark 1:} Even though it cannot be analytically shown that the proposed iterative algorithm converge to a global maximum in \cite{Fesler-94-A}, it is established that, in general, the iterative algorithm monotonically increase the LLF at every iteration and converge to a local maximum. Moreover, if the algorithms are initialized in a region suitably close to the global maximum, then sequence of estimates converge monotonically to the global maximum \cite{Fesler-94-A}. Based on the equivalent system model in \eqref{eq:recsigkantb} and the simulation results in Section \ref{sec:Simresultsdis}, it can be concluded that the proposed iterative algorithm converges globally when the PHN vector $\hat{\boldsymbol{\theta}}_m$ is initialized as $\hat{\boldsymbol{\theta}}_m^{[0]} = [\underbrace{\mathbf{0}_{N-1 \times 1}}_{v=1}, \underbrace{\mathbf{0}_{N-1 \times 1}}_{2},\dots,\underbrace{\mathbf{0}_{N-1 \times 1}}_{N_t}]^T$.     
\section{Complexity analysis} \label{sec:Complexityanalysis}
In this section, the computational complexity of the proposed algorithm is analyzed. Computational complexity is defined as the number of complex additions, $C^{[A]}$, plus number of multiplications, $C^{[M]}$ \cite{Hani:2011Apr}, i.e.,
$C^{[M]}$ and $C^{[A]}$ are determined as
\begin{eqnarray} \label{eq:eqDETMMRD}
C^{[M]}&=& \, N_r\bigg[N_t\Big[\underbrace{(N_t-1)\big[N^2(3N+1)\big]}_{\eqref{eq:SAGE_Estep}}  +\big[\underbrace{N}_{(\ref{eq:eq12})}+\underbrace{5N}_{(\ref{eq:eq15})}+\underbrace{2N}_{(\ref{eq:eq16})}+\underbrace{2N}_{(\ref{eq:eq17})} \nonumber\\
&+&\underbrace{N^2(N+1)}_{\mathbf{s}_v \text{in} (\ref{eq:eq7})}+\underbrace{N^2N_t(NN_t(N_t+2)+1)}_{(\ref{eq:eqLShestforX_1})}+\underbrace{2N^3}_{\boldsymbol{\Gamma}_v \triangleq \textbf{F}^H\mathbf{H}_{v,m}\boldsymbol{\Lambda}_v}\big]t\Big] \nonumber\\
&+&\underbrace{N^2N_t(N N_t(N_t+2)+1)}_{\hat{\mathbf{X}}^{[0]}=(\boldsymbol{\Psi}^{[j-1]H}\boldsymbol{\Psi}^{[j-1]})^{-1}\boldsymbol{\Psi}^{[j-1]H}\mathbf{y}}\bigg],
\end{eqnarray}
\begin{eqnarray} \label{eq:eqDETAMRD}
C^{[A]}&=& \, N_r\bigg[N_t\Big[\underbrace{(N_t-1)\big[N(N-1)(2N+1)\big]}_{\eqref{eq:SAGE_Estep}}  +\big[\underbrace{N}_{(\ref{eq:eq14})}+\underbrace{N}_{(\ref{eq:eq15})}+\underbrace{2N}_{(\ref{eq:eq16})}+\underbrace{N}_{(\ref{eq:eq17})} \nonumber\\
&+&\underbrace{N(N-1)(N+1)}_{\mathbf{s}_v \text{in} (\ref{eq:eq7})}+\underbrace{NN_t(N-1)(NN_t+1)+N^2N_t(NN_t^2+NN_t-1)}_{(\ref{eq:eqLShestforX_1})} \nonumber\\
&+&\underbrace{2N^2(N-1)}_{\boldsymbol{\Gamma}_v \triangleq \textbf{F}^H\mathbf{H}_{v,m}\boldsymbol{\Lambda}_v}\big]t\Big] 
+\underbrace{N N_t(N-1)(N N_t+1)+N^2 N_t(N N_t^2+N N_t-1)}_{\hat{\mathbf{X}}^{[0]}=(\boldsymbol{\Psi}^{[j-1]H}\boldsymbol{\Psi}^{[j-1]})^{-1}\boldsymbol{\Psi}^{[j-1]H}\mathbf{y}}\bigg],
\end{eqnarray}

\begin{figure*}[htbp]
	\begin{minipage}[h]{0.470\textwidth}
		\centering
		\includegraphics[width=1 \textwidth]{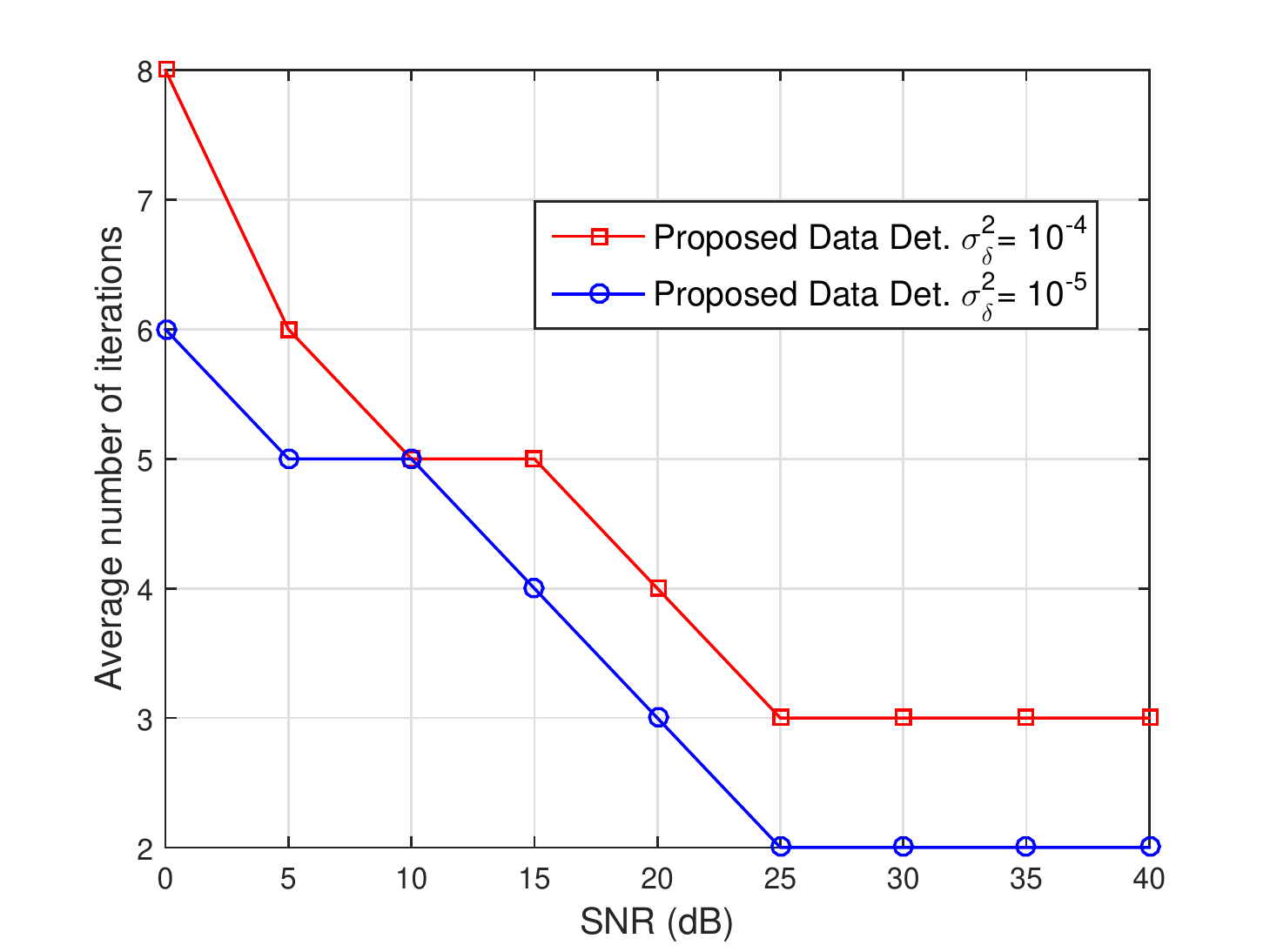}
		\caption{Average number of iterations for the proposed iterative algorithm based on EKF in a $2 \times 2$ MIMO system, phase noise variance $\sigma_\delta^2=[10^{-4},10^{-5}]$ $\text{rad}^2$ and 16-QAM modulation.}
		\label{fig:itersECMEKFinDF}
	\end{minipage}
	\hspace{0.2in}
	\begin{minipage}[h]{0.470\textwidth} \label{fig-12}
		\centering
		\includegraphics[width=1 \textwidth]{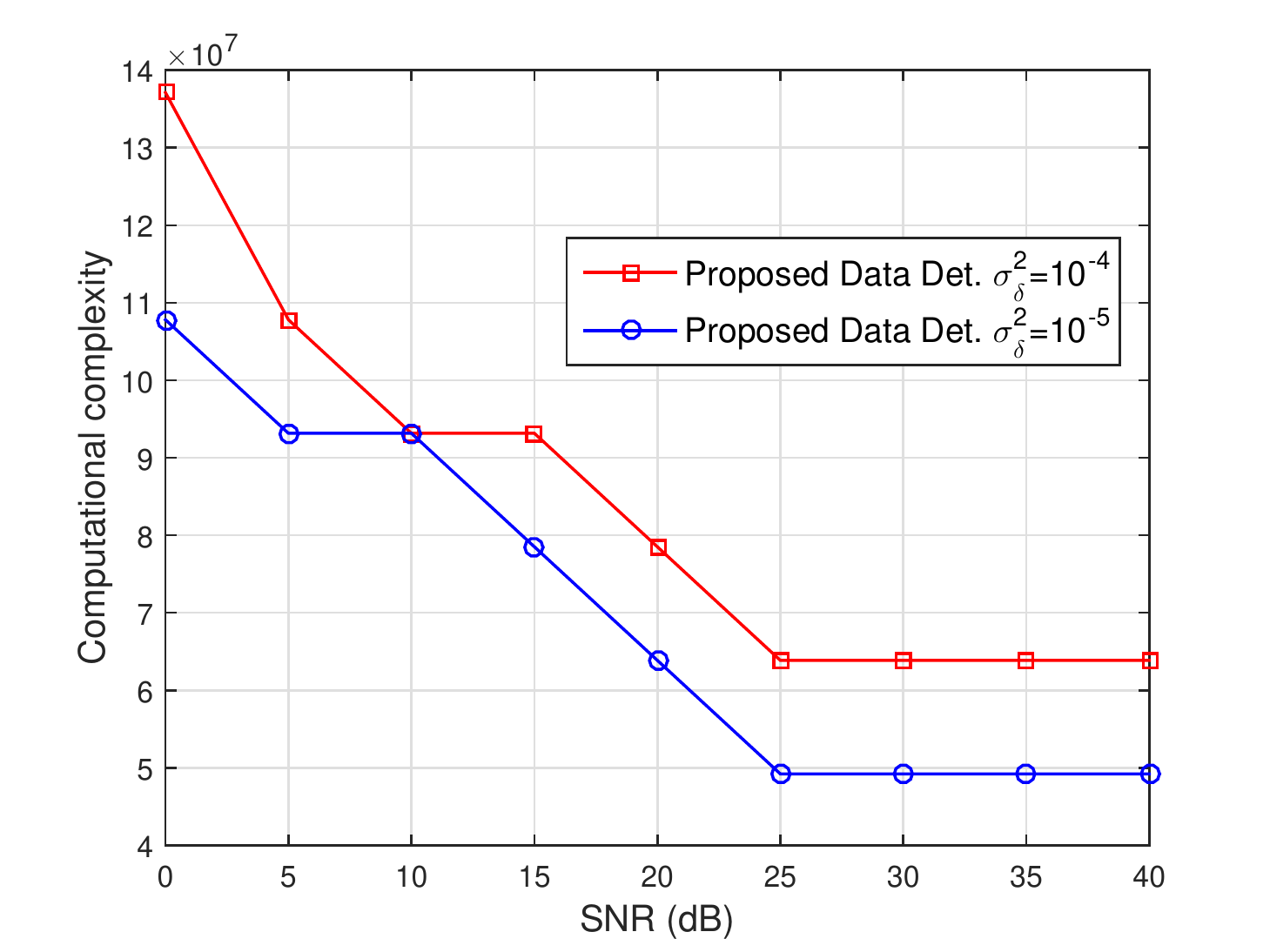}
		\caption{Comparison of the computational complexity of the proposed algorithm in a $2 \times 2$ MIMO system for phase noise variance $\sigma_\delta^2=[10^{-4},10^{-5}]$ $\text{rad}^2$, and 16-QAM modulation.}
		\label{fig:ComputationalcomplexityDF}
	\end{minipage}
\end{figure*}

Fig. \ref{fig:itersECMEKFinDF} shows the average number of iterations, i.e., $t$ required by the proposed algorithm for a $2 \times 2$ MIMO and a PHN variance of $\sigma^2_\delta= [10^{-4}, 10^{-5}] \ \text{rad}^2$ with 16-QAM modulation. It can be observed from the results in Fig. \ref{fig:itersECMEKFinDF} that
(i) at low SNR, i.e., $SNR < 25$ dB, on average, the proposed detector converges after $t$ more than $2$ or $3$ iterations,
(ii) the proposed detector requires more number of iterations at high PHN variance, $\sigma^2_\delta= 10^{-4} \ \text{rad}^2$. This is due to the irreducible error of PHN estimation,
and (iii) the number of iterations decreases to $t=2$ at PHN variance, $\sigma^2_\delta= 10^{-5} \ \text{rad}^2$ and $SNR \geq 25$ dB. Using these values for the number of iterations, we determine the computational complexity of the proposed algorithm for a $2 \times 2$ MIMO system as shown in Fig. \ref{fig:ComputationalcomplexityDF}.
The results in Fig. \ref{fig:ComputationalcomplexityDF} show that (i) at low SNR, i.e., $SNR < 25$ dB, the computational complexity of the proposed algorithm dependent on the variance of the PHN process, since at low SNR the performance of the proposed detector is dominated by AWGN and PHN variance, while at moderate-to-high SNR, i.e., $SNR \geq 25$ dB the performance of the system is limited by residual PHN, and (ii) at moderate-to-high SNR compared to low SNR, the proposed data detection algorithm is computationally more efficient. These results are anticipated, since the proposed data detection algorithm require few iterations, i.e., $2$ or $3$ iterations, at moderate-to-high SNR as shown in Fig. \ref{fig:itersECMEKFinDF}.
\section{Simulation results and discussions} \label{sec:Simresultsdis}
In this section, simulation results are  presented to evaluate the bit error rate (BER) performance of the proposed detection algorithm of the overall MIMO-OFDM system. Moreover, we investigate the effect of joint data detection and estimation of multiple PHN parameters on the PSNR performance of LF video transmission. A multipath Rayleigh fading channel with a delay of $L = 4$ channel taps and an exponentially decreasing power delay profile with the average channel power \mbox{$= [-1.52 -6.75 -11.91 -17.08]$} dB is assumed between each pair of antennas. The sampling rate of the OFDM signal is set to \mbox{$20$} MHz, corresponding to an OFDM sampling duration of \mbox{$50$ nanoseconds}. A training symbol size of $N = 64$ subcarriers is used, where each subcarrier is modulated using quadrature
amplitude modulation (QAM) scheme. The Wiener PHN is generated with different PHN variances, e.g. $\sigma_\delta^2=[10^{-4},10^{-5}] \ \text{rad}^2$. Note that, $\sigma^2_\delta=10^{-4}$ $\text{rad}^2$, corresponds to a high phase noise variance \cite{Hani:2012sept}. In order to evaluate the PSNR performance of the proposed system, several experiments are conducted with a ``Candle" LF video sequence with $378 \times 378$ pixels and a GoP of 10 with 30 frames per second (fps). The captured scene comprises a candle rotating at approximately 3 degrees per frame. A commercial Lytro LF camera was used to capture static LFs of the scene. The rotation centre was about 20 cm from the camera. 120 LF frames were captured under homogeneous ambient illumination so that specular reflections were minimized. Hence, the generated LFV has the size of $10\times 10\times 378\times 378\times 120$.
The video sequence is encoded by H.264 reference software JM version (13.2) in \cite{H264AVCReferenceSoftware:JM}. Following \cite{{Stuhlmuller:2000},{DaweiWang:2013},{DaweiWang:2015}}, the video constants, $a$, $b$, and $z$ of RD curves in \eqref{eq:eqDT}, are calculated using carve fitting. To protect the video data,  a low-density parity-check (LDPC) code is employed with a channel coding rate of $r_c=1/2$ and codeword length of $1296$ bits. The algorithm in \cite{Thomas-2001Feb} is used for encoding. The soft-decision iterative decoding algorithm, based on a sum-product algorithm in \cite{Johnson:Introducing} is utilized for decoding the estimated data vector in \eqref{eq:eqLShestforX_1}. The simulation results are averaged over $1 \times 10^5$ Monte Carlo simulation runs. Finally, in the simulation results, PSNR in dB can be calculated as
\begin{align} \label{eq-PSNRgeneq}
\text{PSNR}=10 \log_{10} \bigg(\frac{255^2}{D_T}\bigg),
\end{align}
where $D_T$ is calculated as shown in \eqref{eq:eqDT}.

\subsection{Impact of PHN} \label{sec-ImPHNoncoopper}
In the following, we examine the performance of the proposed iterative algorithm based on EKF in terms of the BER and PSNR. The following system setups are considered for comparison:

\noindent (i) MIMO systems that applied the proposed data detection algorithm (labelled as ``Proposed Data Det.").

\noindent (ii) The data detection based on pilots in \cite{{Rabiei:2011May},{Wang:arxiv}} (labelled as ``Data Det. based Pilots \cite{{Rabiei:2011May},{Wang:arxiv}}").

\noindent (iii) As a reference, a system that applies the proposed detection algorithm but utilizes no PHNs tracking during OFDM data symbols (labelled as ``No PHNs track.").

\noindent (iv) As a lower- and upper-bound on the BER and PSNR performance, respectively, a system assuming perfect PHNs estimation (labelled ``Perf. PHNs est.").

\noindent (v) To be in line with the assessment of PSNR in the literature, e.g., \cite{LANMAN:2010ACM}, a PSNR of $30$ dB is considered as the lowest PSNR to provide the light field video for the user. 

\begin{figure*}[t]
	\begin{minipage}[h]{0.50\textwidth}
		\centering
		\includegraphics[width=1.05 \textwidth]{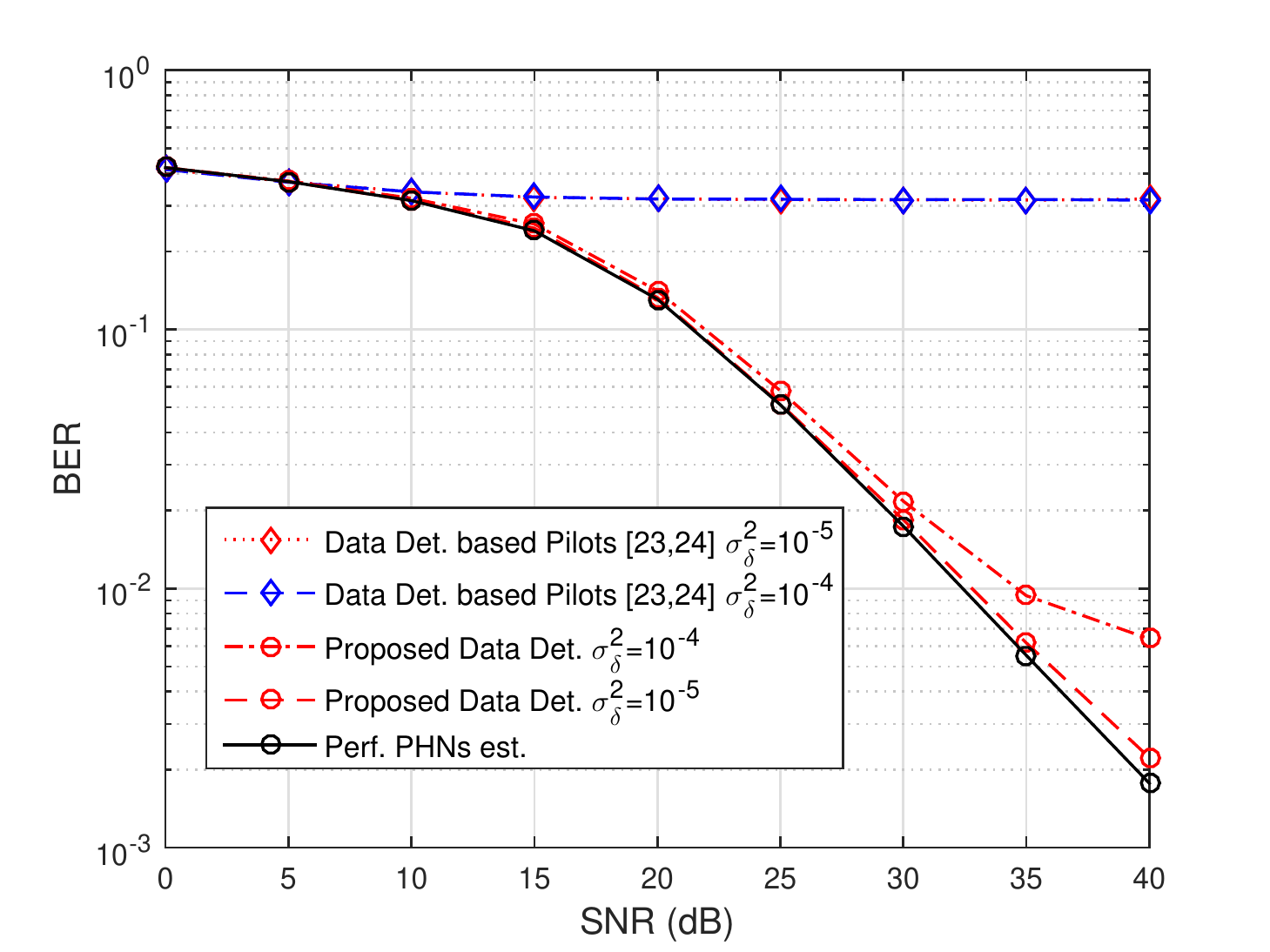}
		\caption{Uncoded BER performance of a $2 \times 2$ MIMO system using the proposed data detector for PHN variance, $\sigma^2_\delta=[10^{-4}, 10^{-5}]$ $\text{rad}^2$ and 16-QAM modulation.}\label{fig:BEReffsPHNs}
	\end{minipage}
	\hspace{0.2in}
	\begin{minipage}[h]{0.50\textwidth}
		\centering
		\includegraphics[width=1.05 \textwidth]{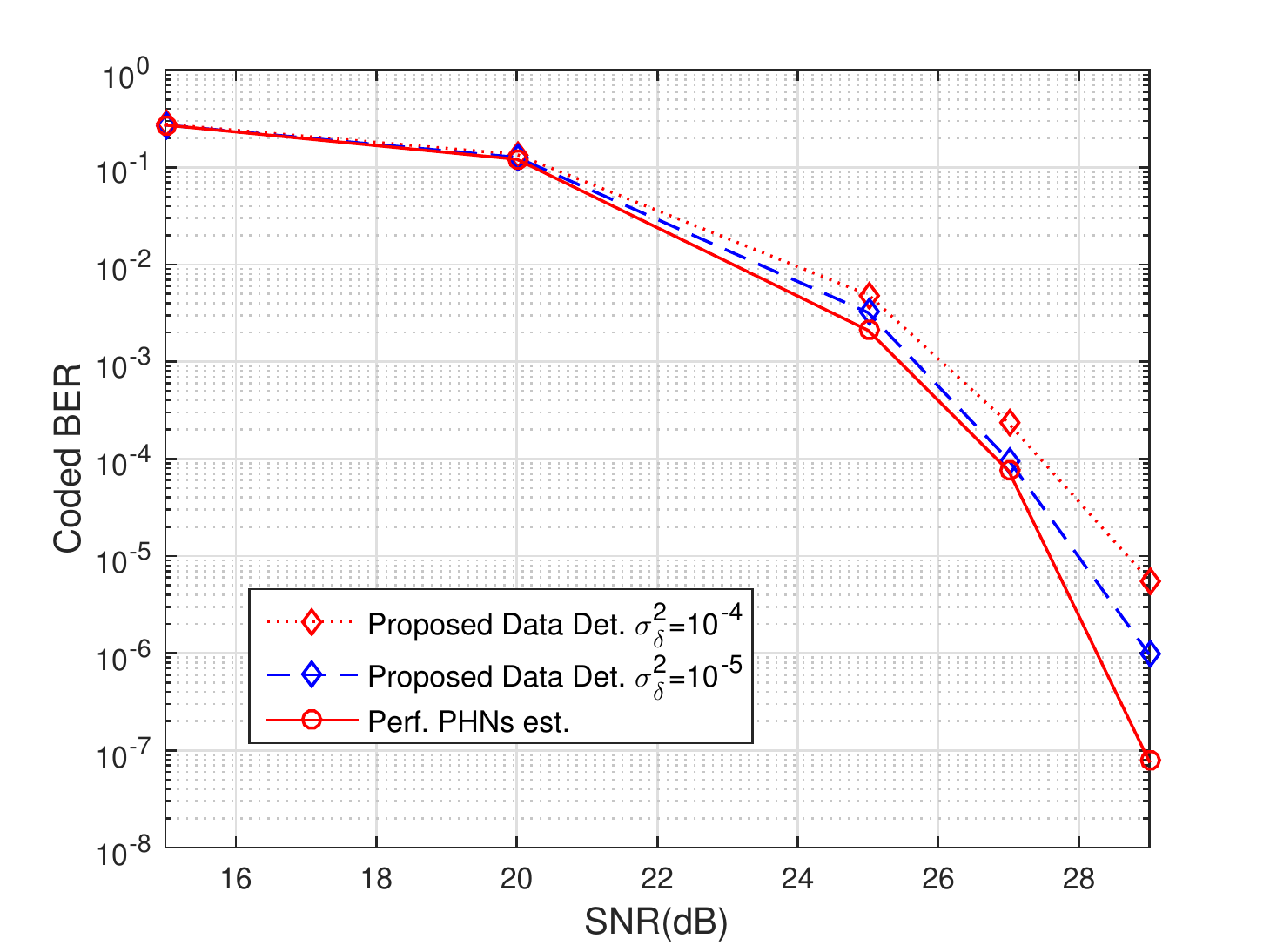}
		\caption{Coded BER performance of a $2 \times 2$ MIMO system using the proposed data detector for PHN variance, $\sigma^2_\delta=[10^{-4}, 10^{-5}]$ $\text{rad}^2$ and 16-QAM modulation.}\label{fig:codedBER}
	\end{minipage}
	\vspace{-0.18in}
\end{figure*}
\begin{figure}
	\begin{center}
		\includegraphics[width=0.53 \textwidth]{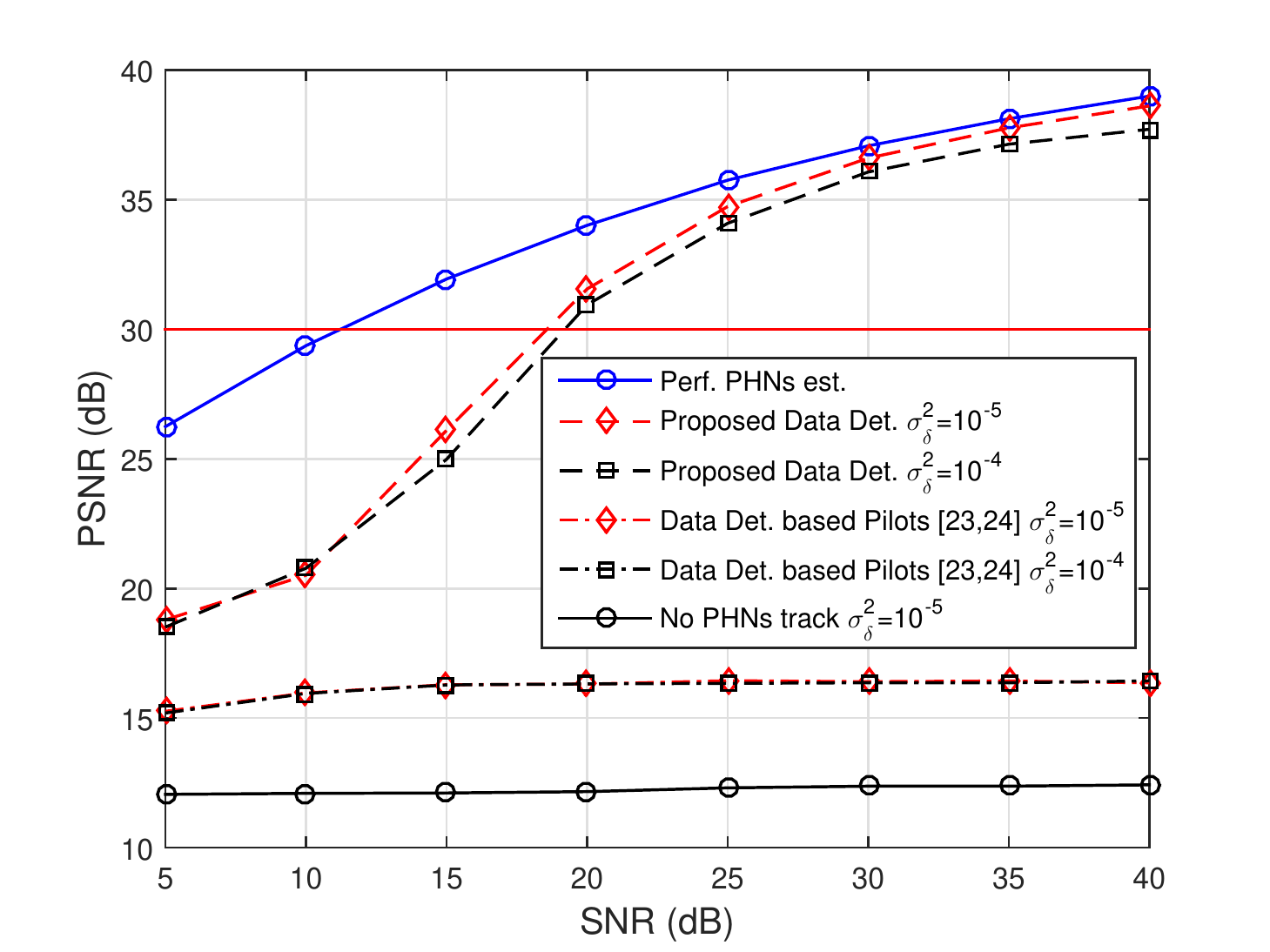}
	\end{center}
	\caption{PSNR performance of a $2 \times 2$ MIMO system using the proposed data detector for PHN variance, $\sigma^2_\delta=[10^{-4}, 10^{-5}]$ $\text{rad}^2$ and 16-QAM modulation.}
	\label{fig:PSNReffsPHNs}
\end{figure}

Figs. \ref{fig:BEReffsPHNs}, \ref{fig:codedBER}, and \ref{fig:PSNReffsPHNs} show the uncoded BER, coded BER, and PSNR performance, respectively, of a $2 \times 2$ MIMO system for PHN variance, $\sigma_\delta^2=[10^{-4},10^{-5}] \ \text{rad}^2$ and 16-QAM modulation.

The following observations can be made from Figs. \ref{fig:BEReffsPHNs}, \ref{fig:codedBER}, and \ref{fig:PSNReffsPHNs}:

\noindent 1) The results in Fig. \ref{fig:PSNReffsPHNs} demonstrate that without PHNs tracking throughout the packet, the PSNR performance of MIMO-OFDM system deteriorates significantly. On the other hand, by combining the proposed data detection algorithm with the MIMO system, the PSNR performance of a MIMO-OFDM system is shown to improve immensely even in the presence of strong PHN, e.g., $\sigma_\delta^2=10^{-4} \ \text{rad}^2$,

\noindent 2) Figs. \ref{fig:BEReffsPHNs}, \ref{fig:codedBER}, and \ref{fig:PSNReffsPHNs} show that the detectors BER and PSNR are dependent on the variance of the PHN process, and are lower and higher, respectively, for a lower PHN variance,

\noindent 3) The BER and PSNR performance using the proposed algorithm significantly outperforms the existing data detection based on pilots in \cite{{Rabiei:2011May},{Wang:arxiv}} at different SNRs. This result is anticipated, since the detection method in \cite{{Rabiei:2011May},{Wang:arxiv}} depends on the pilots, which are effected by the ICI and the interference signals between the antennas. Therefore, the detection approach in \cite{{Rabiei:2011May},{Wang:arxiv}} may not be used for the joint data detection and estimation of multiple PHN parameters.

\noindent 4) The results in Fig. \ref{fig:codedBER} show that, the coded BER performance of a MIMO-OFDM system using the proposed algorithm is close to that of the ideal case of perfect PHNs estimation (a performance gap of $0.8$ dB at BER of $10^{-6}$ and a PHN variance of $10^{-5}$ rad$^2$ ). 

\noindent 5) The results in Fig. \ref{fig:PSNReffsPHNs} show that the PSNR performance of the proposed algorithm is close to that of the ideal case of ``Perf. PHNs est." at moderate-to-high SNR and different PHN variances. For example, as shown in Fig. \ref{fig:PSNReffsPHNs}, at SNR = 25 dB, the PSNR loss of the proposed system are almost 1.6 dB and 1 dB for PHN variances $\sigma_\delta^2=[10^{-4},10^{-5}] \ \text{rad}^2$, respectively, compared to the case of ``Perf. PHNs est.".

\noindent 6) The results in Fig. \ref{fig:PSNReffsPHNs} show that the proposed algorithm significant improves the PSNR at different PHN variances and is capable to provide light field video for more than $30$ dB at moderate-to-high SNR, i.e., $SNR \geq 23$ dB.

\noindent 7) Compared to the ``Perf. PHNs est.", at high PHN variance, i.e., $\sigma_\delta^2=10^{-4} \ \text{rad}^2$, the BER and PSNR performance suffer from an error floor at high SNR. This result is anticipated, since at high PHN variance, the performance of a MIMO-OFDM system is dominated by PHN, which cannot be completely eliminated.

\subsection{Impact of modulation} \label{sec-Immodoncoopper}
\begin{figure*}[t]
	\begin{minipage}[h]{0.50\textwidth}
		\centering
		\includegraphics[width=1.05 \textwidth]{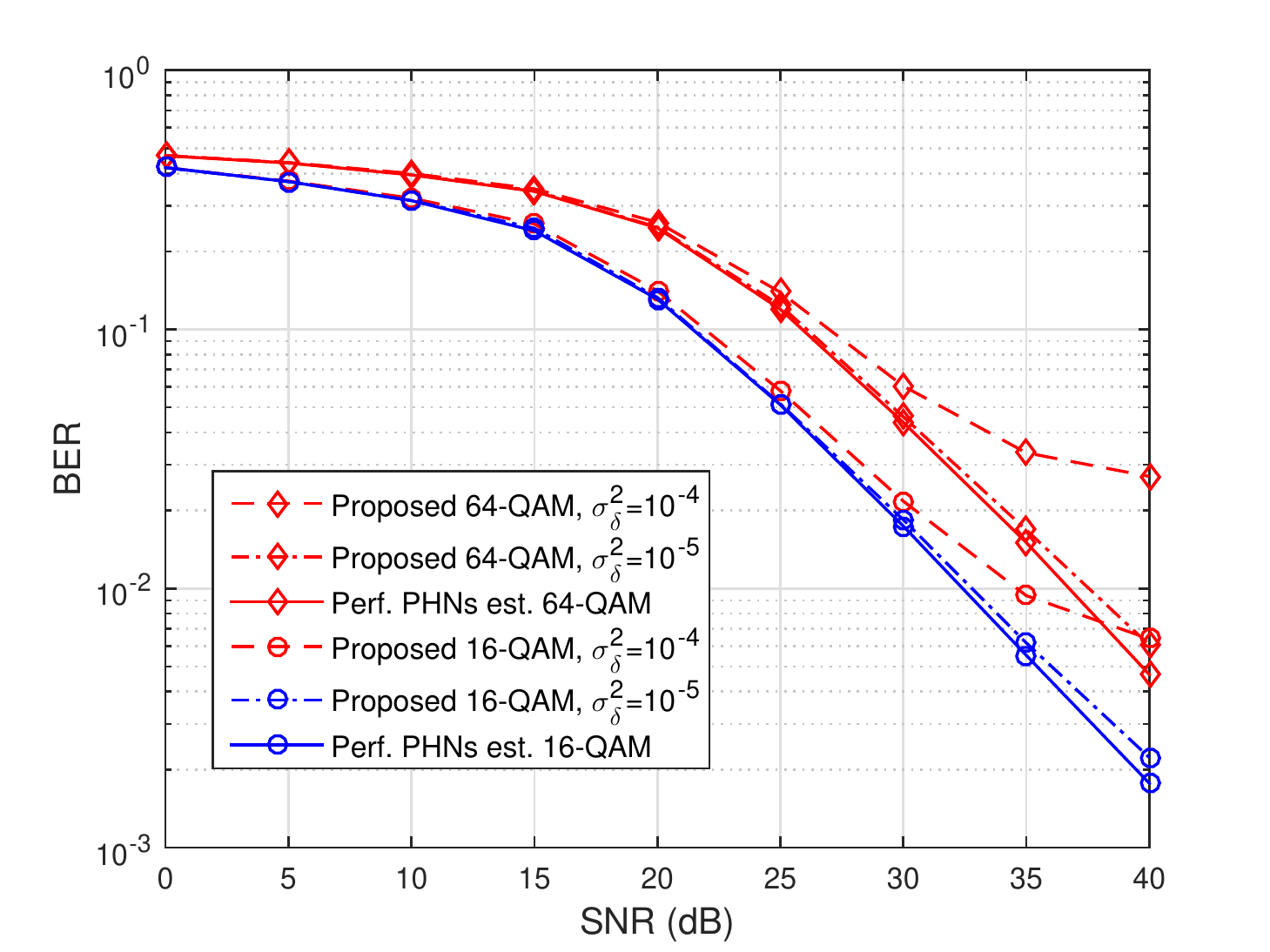}
		\caption{BER performance of a $2 \times 2$ MIMO system using the proposed data detector for PHN variance, $\sigma^2_\delta=[10^{-4}, 10^{-5}]$ $\text{rad}^2$ and 64-QAM modulation.}\label{fig:BEReffectsmods}
	\end{minipage}
	\hspace{0.2in}
	\begin{minipage}[h]{0.50\textwidth}
		\centering
		\includegraphics[width=1.05 \textwidth]{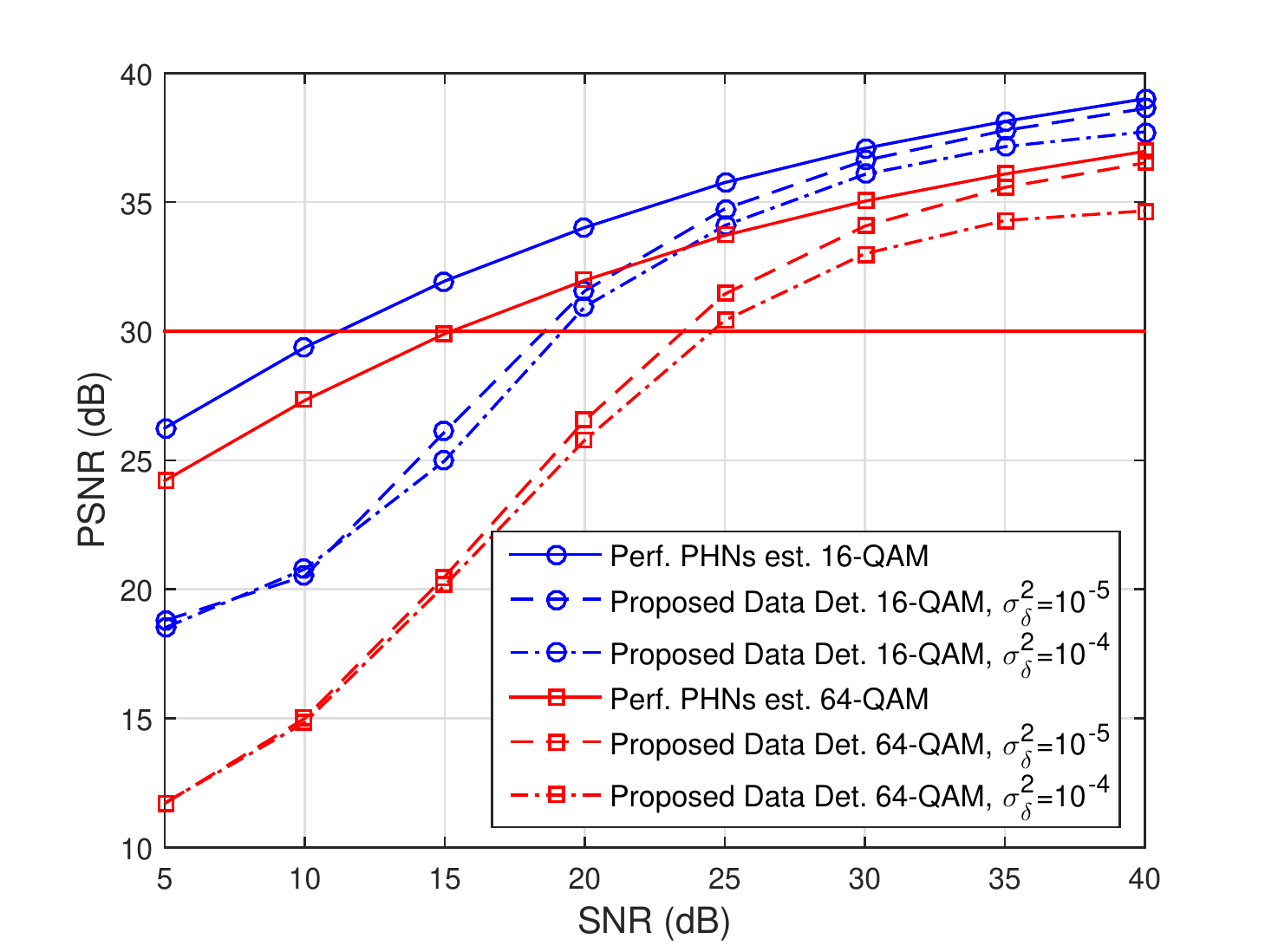}
		\caption{PSNR performance of a $2 \times 2$ MIMO system using the proposed data detector for PHN variance, $\sigma^2_\delta=[10^{-4}, 10^{-5}]$ $\text{rad}^2$ and 64-QAM modulation.}\label{fig:PSNReffectsmods}
	\end{minipage}
	\vspace{-0.18in}
\end{figure*}

Figs. \ref{fig:BEReffectsmods} and \ref{fig:PSNReffectsmods} evaluate the BER and PSNR performance of the MIMO system at higher order modulations, i.e., 64-QAM.

The following observations can be made from Figs. \ref{fig:BEReffectsmods} and \ref{fig:PSNReffectsmods}:

\noindent 1) Even for denser constellation, the proposed data detection algorithm achieve BER and PSNR performance that are closer to the ideal case of perfect PHNs estimation. For example, as shown in Fig. \ref{fig:PSNReffectsmods}, at SNR of $25$ dB and a PHN variance of $10^{-4}$ $\text{rad}^2$ and 64-QAM, the PSNR performance of MIMO system is close with 3.2 dB to the ideal case of perfect PHNs estimation.


\noindent 2) The results in Figs. \ref{fig:BEReffectsmods} and \ref{fig:PSNReffectsmods} show that the PHN variance and the order of modulation play significant factors in determining the BER and PSNR performance. For instance, as shown in Fig. \ref{fig:PSNReffectsmods}, at SNR of $20$ dB and a PHN variance of $10^{-4}$ $\text{rad}^2$ and 64-QAM, the system can achieve PSNR gain of $5.17$ dB by lower orders of modulation, e.g. 16-QAM.

\noindent 3) Even for denser constellation, the results in Figs. \ref{fig:PSNReffectsmods} show that the proposed data detection achieve a PSNR more than $30$ dB at $SNR \geq 23$ dB at different PHN variances, i.e., SNR loss of $5$ dB compared to 16-QAM.

\subsection{Impact of increase number antennas} \label{sec-ImPHNoncoopper}
\begin{figure*}[t]
	\begin{minipage}[h]{0.50\textwidth}
		\centering
		\includegraphics[width=1.05 \textwidth]{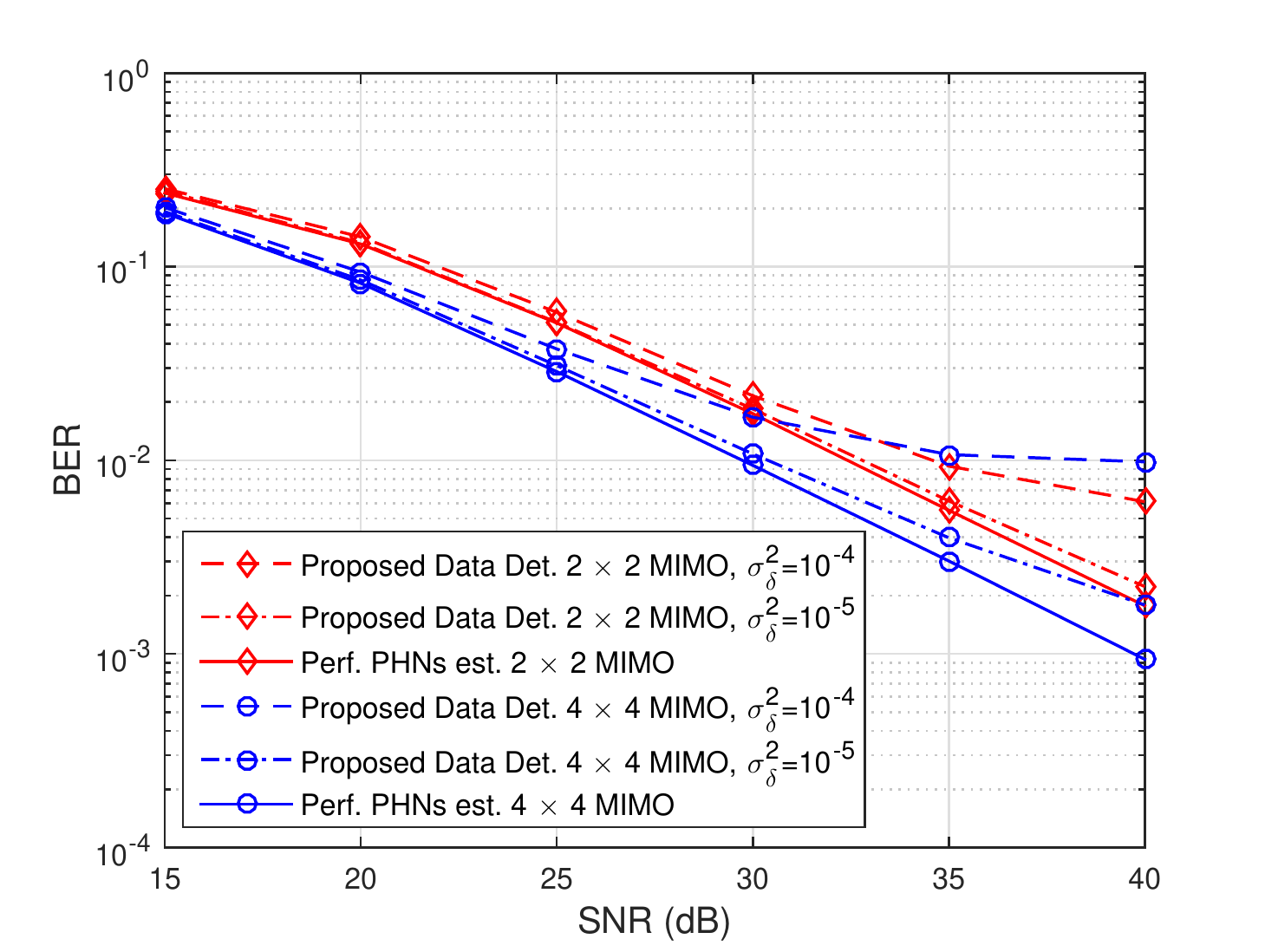}
		\caption{BER performance of a $4 \times 4$ MIMO system using the proposed data detector for PHN variance, $\sigma^2_\delta=[10^{-4}, 10^{-5}]$ $\text{rad}^2$ and 16-QAM modulation.}\label{fig:BEReffectsants}
	\end{minipage}
	\hspace{0.2in}
	\begin{minipage}[h]{0.50\textwidth}
		\centering
		\includegraphics[width=1.05 \textwidth]{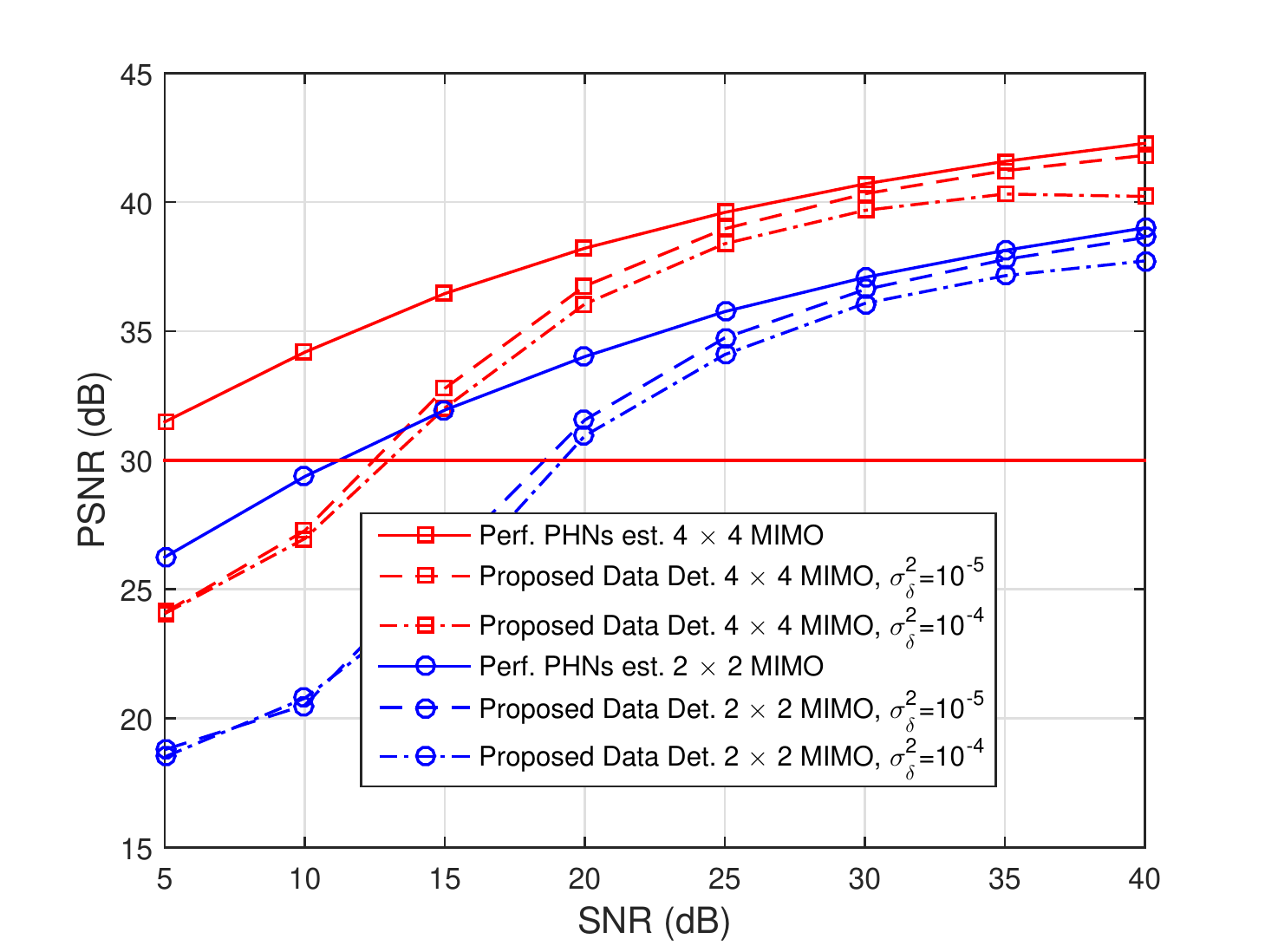}
		\caption{PSNR performance of a $4 \times 4$ MIMO system using the proposed data detector for PHN variance, $\sigma^2_\delta=[10^{-4}, 10^{-5}]$ $\text{rad}^2$ and 16-QAM modulation.}\label{fig:PSNReffectsants}
	\end{minipage}
	\vspace{-0.18in}
\end{figure*}

Figs. \ref{fig:BEReffectsants} and \ref{fig:PSNReffectsants} show the BER and PSNR performance at different number of antennas, $N_t=N_r=[2, 4]$ for PHN variance, $\sigma_\delta^2=[10^{-4},10^{-5}] \ \text{rad}^2$ and 16-QAM modulation. The following observations can be made from Figs. \ref{fig:BEReffectsants} and \ref{fig:PSNReffectsants}:

\noindent 1) At moderate PHN variance, i.e., $\sigma_\delta^2=10^{-5} \ \text{rad}^2$, although the increase number of antennas leads to increase the interference and ICI signals, the BER and PSNR performance are improved and closer to the ideal case of perfect PHNs estimation. For example, in Figs. \ref{fig:BEReffectsants}, at $\sigma_\delta^2=10^{-5} \ \text{rad}^2$, a performance gap of 0.3 dB and 0.2 dB at BER = $10^{-2}$ for a $2 \times 2$ and $4 \times 4$ MIMO, respectively.

\noindent 2) At high PHN variance, i.e., $\sigma_\delta^2=10^{-4} \ \text{rad}^2$, the BER and PSNR performance degrades with increase of number of antennas. This result is anticipated, since at high PHN variance, the performance of proposed detector is impacted by the considerable residual phase noise estimation error. Therefore, in the presence of high PHN variance, the MIMO system can achieve significant BER and PSNR performance by combining the proposed data detection algorithm and using few antennas. However, this approach maintains higher performance at the expense of loss in spectral efficiency.

\noindent 3) At high SNR regime and high PHN variance in , i.e., $\sigma_\delta^2=10^{-4} \ \text{rad}^2$, the BER and PSNR performance suffer from an error floor. This is due to the fact that at low SNR the performance of the system is dominated by AWGN, while at high SNR the performance of the proposed detector is limited by PHN and the resulting ICI.

\noindent 4) The results in Fig. \ref{fig:PSNReffectsants} for a $4 \times 4$ MIMO system show the proposed detector achieves a PSNR performance more than $30$ dB in the low SNR regime, i.e., SNR$=12$ dB, at PHN variance $\sigma_\delta^2=10^{-5} \ \text{rad}^2$. This indicates that the proposed detector can achieve high PSNR at different PHN variances.
 
\section{Conclusion} \label{sec:sec6}
In this paper, we addressed the joint data detection and phase noise mitigation of multiple PHN parameters for LF video transmission in MIMO-OFDM systems. An iterative algorithm for joint data detection and PHN mitigation was proposed for MIMO-OFDM data symbols. We also investigated the improvement in PSNR for LF video transmission by combining the proposed algorithm and MIMO-OFDM systems in the present of multiple PHN parameters. The proposed detector was found to be computationally efficient, which can detect the desired data parameters in a few iterations. Numerical results showed that the proposed detector can significantly improve the average BER and PSNR performance for LF video transmission compared to existing algorithms. Moreover, the BER and PSNR performance of the proposed system were closer to the ideal case of perfect PHNs estimation. It was demonstrated that the proposed system model and algorithm are well suited for LF video transmission in wireless channels. 

\vspace{6pt}

\ifCLASSOPTIONpeerreview
\vspace{-0.5cm}
\else
\fi

\end{document}